\documentclass[a4paper,fleqn,usenatbib]{mnras}

\usepackage{newtxtext,newtxmath}

\usepackage[T1]{fontenc}
\usepackage{ae,aecompl}
\usepackage{booktabs,caption}
\usepackage[flushleft]{threeparttable}

\usepackage[nolist,nohyperlinks,printonlyused]{acronym}

\usepackage{float}

\usepackage{graphicx}	
\usepackage{amsmath}	
\usepackage{comment}

\newcommand{\HI}{\ion{H}{I}}

\newcommand{\mathHI}{{\mbox{\scriptsize \HI}}}

\newcommand\given[1][]{\:#1\vert\:}

\newcommand{\lyaf}{\text{Ly$\alpha$ forest}}
\newcommand{\lya}{\text{Ly$\alpha$}}
\newcommand{\lyb}{\text{Ly$\beta$}}
\newcommand{\NHI}{$N_{\mathHI{}}$}

\newcommand{\NHIt}{N_{\mathHI{}} }

\newcommand{\vpfit}{\texttt{VPFIT}}
\newcommand{\bndist}{$b$-$N_{\mathHI{}}$~distribution}
\newcommand{\bn}{$\left\{b, N_{\mathHI{}}\right\}$}
\newcommand{\dndz}{d$N$/d$z$}

\defcitealias{Hu2022}{Hu22}

\title[The IGM Thermal and Ionization State at $z \sim 1$]{Measurements of the Thermal and Ionization State of the Intergalactic Medium during the Cosmic Afternoon}

\author[Hu et al.]{Teng Hu,$^{1}$\thanks{E-mail: tenghu@ucsb.edu (UCSB)}
Vikram Khaire$^{1,2}$,
Joseph F. Hennawi$^{1,3}$,
Todd M. Tripp$^{4}$,
Jose O\~norbe$^{5}$, \newauthor
Michael Walther$^{6,7}$, and
Zarija Lukic$^{8}$ 
\\
$^{1}$Physics Department, Broida Hall, University of California Santa Barbara, Santa Barbara, CA
93106-9530, USA\\
$^{2}$ Indian Institute of Space Science \& Technology, Thiruvananthapuram, Kerala - 695547, INDIA\\
$^{3}$Leiden Observatory, Leiden University, PO Box 9513, NL-2300 RA Leiden, the Netherlands\\
$^{4}$Department of Astronomy, University of Massachusetts, Amherst, MA 01003, USA\\
$^{5}$Facultad de F\'isica, Universidad de Sevilla, Avda. Reina Mercedes s/n, Campus de Reina Mercedes, E-41012 Sevilla, Spain\\
$^{6}$University Observatory, Faculty of Physics, Ludwig-Maximilians-Universität München, Scheinerstr. 1, 81679 Munich, Germany\\
$^{7}$Excellence Cluster ORIGINS, Boltzmannstr. 2, 85748 Garching, Germany\\
$^{8}$Lawrence Berkeley National Laboratory, Berkeley, CA 94720, USA\\
}

\date{Accepted XXX. Received YYY; in original form ZZZ}

\pubyear{2015}

\begin{document}
\label{firstpage}
\pagerange{\pageref{firstpage}--\pageref{lastpage}}
\maketitle

\begin{abstract}

We perform the first measurement of the thermal and ionization state of the intergalactic medium (IGM) across $0.9 < z < 1.5$ using 301 \lya{} absorption lines fitted from 12 archival HST STIS quasar spectra, with a total pathlength of $\Delta z=2.1$.
We employ the machine-learning-based inference method that uses joint Doppler parameter - column density ($b-N_{\mathHI{}}$) distributions 
obtained from \lyaf{} decomposition.
Our results show that the H~{\sc i} photoionization rates,
$\Gamma_{\mathHI{}}$, are in good agreement with the recent UV background synthesis models, 
with $\log (\Gamma_{\mathHI}/\text{s}^{-1})={-11.79}^{+0.18}_{-0.15}$, ${-11.98}^{+0.09}_{-0.09}$, 
and ${-12.32}^{+0.10}_{-0.12}$ at $z=1.4$, $1.2$, and $1$ respectively.
We obtain the IGM temperature at the mean density, $T_0$, and the adiabatic index, $\gamma$,
as $[\log (T_0/\text{K}), \gamma]=$ [${4.13}^{+0.12}_{-0.10}$, ${1.34}^{+0.10}_{-0.15}$], $[{3.79}^{+0.11}_{-0.11}$, ${1.70}^{+0.09}_{-0.09}]$ and $[{4.12}^{+0.15}_{-0.25}$, ${1.34}^{+0.21}_{-0.26}]$  at $z=1.4$, $1.2$ and $1$ respectively.
Our measurements of $T_0$ at $z=1.4$ and $1.2$ are consistent with the expected trend from $z<3$ temperature measurements as well as theoretical expectations that, in the absence of any non-standard heating, the IGM should cool down after He~{\sc ii} reionization.
Whereas, our $T_0$ measurements at $z=1$ show unexpectedly high IGM temperature. 
However, because of the relatively large uncertainty in these measurements of the order of 
$\Delta (T_0) \sim 5000$ K, mostly emanating from the limited redshift path length 
of available data in these bins, we can not definitively conclude whether the IGM cools down at $z < 1.5$.
Lastly, we generate a mock dataset to test the constraining power of future measurement with larger datasets.
The results demonstrate that, with redshift pathlength $\Delta z \sim 2$ for each redshift bin, 
three times the current dataset, we can constrain the $T_0$ of IGM within $1500$K. 
Such precision would be sufficient to conclusively constrain the history of IGM thermal evolution at $z < 1.5$.


\end{abstract}




\section{Introduction}
\label{sec:intro}

After hydrogen reionization ($z < 6$) \citep{Madau1998,Fan2006,Faucher-Giguere2008,Robertson2015,mcgreer1}, 
the thermal state of the intergalactic medium (IGM) is determined by the balance between heating from photoionization by the extragalactic UV background (UVB) and cooling mechanisms, including adiabatic cooling because of the Hubble expansion, radiative recombination cooling, and inverse Compton scattering where electrons interact with the cosmic microwave background.
As a result of these processes, after the epoch of reionization, the \ac{IGM} subsequently adheres to the power-law temperature-density ($T$-$\Delta$) relation: 
\begin{equation}
T(\Delta) = T_0 \Delta^{ \gamma -1},
\label{eqn:rho_T}
\end{equation}
where $\Delta = \rho/ \bar{\rho}$ is the overdensity, $T_0$ is the temperature at mean density $\bar{\rho}$,
and $\gamma$ is the power-law index \citep{hui1, McQuinn2016}. 
These two parameters $[T_0,\gamma]$ thus characterize the thermal state of the \ac{IGM}, 
and enable us to impose constraints on its thermal history at various epochs \citep{Dave&Tripp2001, Becker2011, Rorai2017, Hiss2018, Gaikwad2021},
which enhance our understanding of the IGM thermal evolution and illustrate the intrinsic heating and cooling mechanisms of the Universe.

Based on current theoretical models, by the later stages of the universe ($z\lesssim 1.7$), 
i.e. long after the end of helium reionization at $z \sim 3$ \citep{mcquinn09,Worseck2011,Khaire2017},
the thermal state of the IGM is dominated by the adiabatic cooling driven by Hubble expansion.
Consequently, it is predicted that the IGM cools to temperatures around $T_0\sim 5000$ K and $\gamma \sim 1.6$ by $z\sim 0$ \citep{McQuinn2016}. 
Interestingly, the specifics of He~{\sc ii} reionization hardly influence this outcome \citep{Onorbe2017,Onorbe2017b}. 
This is because, after roughly 200 Myr, 
the IGM essentially "forgets" its past thermal history due to the aforementioned adiabatic cooling. Standard hydrodynamical simulations routinely forecast this cooling pattern of the IGM to temperatures of $T_0\sim 5000$ K by $z\sim 0$, however,
there is not enough observational evidence to confirm this claim. 
This is mainly because of the large scatter in the $T_0$ measurements performed using 
various techniques over the last two decades at $1.7 < z < 3$, 
although most recent measurement \citep{Hiss2018, Walther2019, Gaikwad2021} 
hint towards cooling down of the IGM at $z<3$. 
The most conclusive evidence of the cooling of the IGM should come from the $T_0$ measurements at $z<1.5$ however these measurements are challenging. 
Part of the challenge lies in the fact that for $z \lesssim 1.7$, the \lya{} transition is obstructed by the atmospheric cutoff ($\lambda \sim 3300$ \AA{}), necessitating UV space observations via Hubble Space Telescope (HST), 
the only space-based telescope that has FUV and NUV spectrographs capable of providing data for these measurements. 
Currently, the only measurements of the IGM thermal state for $z < 1.7$ come from \citet{ricotti2000} and 
\citet{Dave&Tripp2001}, 
both utilizing 
datasets with very limited size ($\sim$ 50 \lya{} absorption lines).
Due to the limited scale of this dataset, 
the associated error margins are substantial, with $\sigma_{T_0} \gtrsim$ 5000K.
This limitation implies that our understanding of the IGM therm state at low-$z$ remains imprecise.

Recent analyses of \ac{HST} \ac{COS} \lya{} absorption spectra at $z < 0.5$ suggest that the IGM temperature at low-$z$ may exceed theoretical predictions \citep{Gaikwad2017, Viel2017, Nasir2017}. 
This claim stems from the decomposition of \lya{} lines, 
where each line is characterized by its Doppler parameter $b$ and the neutral hydrogen column density $N_{\rm HI}$. 
These studies show that the low-$z$ \lya{} lines appear notably broader than anticipated, 
indicated by larger $b$-parameters compared with those obtained from hydro simulations
with and without feedback \citep{Bolton2021, Hu2023, Khaire2023}. 
Since these hydro simulations model the low-density gas traced by the \lya{} forest, from the first principles,
the most straightforward interpretation for these enlarged $b$ values is thermal broadening, suggesting an unexpectedly high IGM temperature \citep{Viel2017} 
or non-standard missing turbulence in the simulations \citep{Gaikwad2017, Bolton2021}.
If the IGM temperature is indeed higher than current models predict, 
it necessitates a reconsideration of heating sources. 
Potential mechanisms might include feedback effects from galaxy formation processes \citep[however see][]{Khaire2023, Khaire2023_halos, Hu2023}
or more novel phenomena such as heating due to dark matter annihilation \citep{Bolton22}. Additionally, turbulent broadening from non-gravitational forces, which are not yet integrated into simulations, could also play a role \citep{Gaikwad2017,Bolton2021}.
It is noteworthy that this discrepancy in $b$ parameter distributions is found only in low-$z$. In contrast, at $z \gtrsim 2$, the distribution of $b$ parameters aligns well with the predictions made by hydro simulations regarding thermal and turbulent broadening \citep[e.g][]{bolton2014, Hiss2019}.

In addition to the discrepancy in $b$ parameters, 
the low-$z$ IGM presents another puzzle: the nature of the UV background (UVB), 
characterized by the H~{\sc i} photoionization rate, $\Gamma_{\mathHI{}}$, 
which directly affects the abundance of \lya{} absorbers in the low-$z$ IGM
as well as crucial for studying the circum-galactic medium \citep[e.g][]{Lehner_2013,Hussain17,Chen17,Wotta_2019,Acharya2022}.
A notable deviation between the $\Gamma_{\mathHI{}}$ deduced from the \lyaf{} at $z\sim 0.1$ and the forecasts from previous UVB synthesis models \citep[e.g][]{H&M2012, Faucher-Gigu2009} lead \citet{Kollmeier2014} to introduce the problem of a "photon under-production crisis",  which has however not been confirmed by other studies \citep[]{Shull2015, Gaikwad2017b, Fumagalli17, Khaire2019} 
and recent UVB models \citep{Khaire15puc, Khaire_Srianand2019, Puchwein19, FG2020}. 
The recent UVB models agree to the extent that the low-z $\Gamma_{\mathHI{}}$ measurements 
favour UVB dominated by H~{\sc i} 
ionizing photons from quasars alone and the fraction of ionizing photons 
from galaxies at $z<2$ is negligibly small \citep{Khaire_Srianand2019, Puchwein19, FG2020}. 
However, at higher redshifts, $z>3$,
a substantial increase in the ionizing escape fraction from galaxies from less than one percent to 15-20 percent is needed \citep{Khaire2016} even in the presence of a high fraction of low-luminosity quasars claimed to be present at high-$z$ \citep{Khaire2017, Finkelstein2019}.
This transition of escape fraction hinges only on the $\Gamma_{\mathHI{}}$ measurements
at $z > 2$ and $z<0.5$ whereas there are no measurement of $\Gamma_{\mathHI{}}$
at $0.5<z<1.8$, with a substantial void of almost five billion years of cosmic time. 
A part of this lack of measurement, 
besides the limited data from HST at these redshifts,
is caused by the potential degeneracy between the IGM thermal and ionization states.  
To overcome this for $z<0.5$ $\Gamma_{\mathHI{}}$ measurements previous studies \citep{Gaikwad2017b, Khaire2019} leveraged 
either post-processing simulations to generate the thermal histories \citep{Gaikwad2018}
or a huge grid of NyX simulations \citep{Walther2017} performed with different thermal histories of the IGM. 
It is important to recognize, a full description of the \lyaf{} depends on three 
parameters $T_0$, $\gamma$, and $\Gamma_{\mathHI{}}$.  
Degeneracies among these variables require that any reliable data-model comparison must adopt a careful statistical inference procedure. 

To overcome the aforementioned difficulties, 
\citet[][hereafter \citetalias{Hu2022}]{Hu2022} adopts an inference method 
which jointly measures the thermal and ionization state of the low-$z$ IGM based on the decomposition of the \lya{} forest into Doppler broadening parameter $b$ and column density $N_{\mathHI{}}$.
In this framework, Bayesian inference of the model parameters 
[$\log T_0$, $\gamma$, $\log \Gamma_{\mathHI{}}$] is conducted based on the
2D joint \bndist{} and the line density \dndz{}, with the help of neural density estimators \citep{Alsing2019} and Gaussian emulators \citep{Ambikasaran2016}, both trained on a suite of Nyx simulations consisting of 51 simulation models with different thermal histories \citep{Walther2017,Hiss2018}. 
Such an inference method enables us to measure the thermal and ionization state of the IGM to high precision using limited-sized data.

In this work, we employ the aforementioned method to measure
both the thermal and ionization state of the IGM using quasar spectra obtained from \ac{STIS} on board \ac{HST}.
We opt for STIS due to its superior resolution compared with \ac{COS} and available archival data. 
We utilize 12 \ac{HST} \ac{STIS} quasar spectra covering $0.9 < z < 1.5$,
which are selected from the STIS archive based on their redshift coverage, \ac{SNR}s, 
and the availability of metal identification.
For the identification of metal lines, we import the metal identification from the \ac{CASBaH} project \citep{CASBaH_data,CASBaH_OVI,CASBaH_NeVIII,CASBaH2021} for five of our spectra,
and make use of the metal identification from \citet{Milutinovic2007} for the remaining seven spectra.
We fit these spectra to obtain our \bn{} sample using \vpfit{} (see \S~\ref{sec:vpfit}) and 
apply the \citetalias{Hu2022} method to measure the thermal and ionization state of the IGM in three redshift bins centering on $z=1$, 1.2, and 1.4.

This paper is structured as follows. 
We introduce our observational data in \S~\ref{sec:data} together with the data processing procedure, 
including continuum fitting, Voigt profiles fitting, 
and metal masking.
In \S\ref{sec:simulations} we describe our hydrodynamic simulations,
parameter grid, and mock data processing procedures, including generating \lyaf{} from simulation,
creating mock sightlines, 
and forward-modeling. 
In \S\ref{sec:inference} we present our inference algorithm, 
including emulators and likelihood function.
Afterwards, we discuss our results in \S\ref{sec:discussion}. 
In the end, we summarize the highlights of this study in \S\ref{sec:conclusion}.
Throughout this paper, we write  $\log$ in place of $\log_{10}$.
Cosmology parameters used in this study 
($\Omega_m = 0.319181, \Omega_b h^2 = 0.022312, h= 0.670386, n_s = 0.96, \sigma_8 = 0.8288$) are taken from \citet{Planck2014}.

\section{Observational Data}
\label{sec:data}

 \begin{figure*}
\centering
    \includegraphics[width=0.95\linewidth]{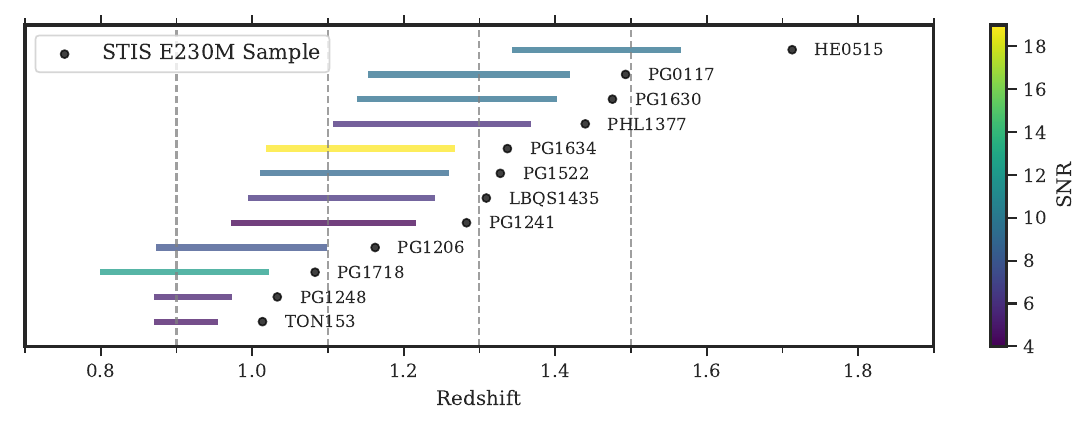}
  \caption{ The HST STIS E230 spectra used in this study. The quasar are shown as black dots,
  and the \lya{} spectra, with proximity zones removed, are shown as line segments with their colour indicating the mean SNR (per pixel). 
  The three redshift bins used in this study are shown by the vertical dashed lines. }
  \label{fig:STIS_spectra}
\end{figure*} 

To measure the thermal state of the IGM around z $\sim 1$,
we make use of the quasar spectra observed with the \ac{HST} \ac{STIS} \citep{Woodgate1998} using the E230M echelle mode,  
which provide spectroscopic coverage from $\sim$ 1600~\AA{} to 3100 \AA.
We select such echelle mode for two reasons. 
First, as discussed in \S~\ref{sec:intro}, 
its high spectral resolution is beneficial for our analysis,
with R $\sim $ 30,000, corresponding to $\sim 10 \text{ km/s}$ \citep{Kimble1998,Medallon2023}, 
and its \ac{LSF} is close to Gaussian and has a weak dependence on the wavelength, 
which makes both the Voigt profile fitting (see \S~\ref{sec:vpfit}) and the generation of forward models easier (see \S~\ref{sec:FM}).
Secondly, the echelle modes have higher wavelength coverage compared with first-order grating modes,
enabling us to measure the \bn{} of the \lya{} absorption lines across a wider redshift range with constant instrumental effects such as \ac{LSF}, which makes our analysis across different redshift bins more robust. 
We search the archival HST STIS E230M data observed in the 0.2'' × 0.2'' slit,
and retrieve 12 spectra with average SNR $\gtrsim 5$. 
The details of the observation, from which our quasar samples are obtained, are summarized in Table~\ref{tab_spec}, and Fig.~\ref{fig:STIS_spectra} depicts the redshift coverage of the spectra used in this study. The quasars are shown as black dots, and the spectra are shown as line segments with their colour indicating the SNR. 
The redshift bins considered for the measurements are shown by the vertical dashed lines in  Fig.~\ref{fig:STIS_spectra}.

To reduce and combine the STIS spectra, we used the procedure of \citet{Tripp2001} with \texttt{CALSTIS v3.4.2}. In brief, starting with the \texttt{CALSTIS x1d} files, for each quasar we combined all exposures, including the coaddition of overlapping regions of adjacent echelle orders, all with appropriate weighting and using the STIS flags to mask out bad pixels \citep[see][for details]{Tripp2001}. 
We then fit the continuum of these spectra using the interactive continuum fitting program imported from {\tt linetools}\footnote{For more information, visit https://linetools.readthedocs.io}. 
Since we focus on the \lyaf{} in this study, we make use of only the \lya{} regions, excluding Ly$\beta$ and higher Lyman series absorption lines at $\lambda < 1050$~\AA, while also masking the quasar proximity zones at $\lambda > 1180$~\AA~(see Fig.\ref{fig:STIS_spectra}). 
As a result, we only use the spectral segment with rest frame wavelength $ 1050 <\lambda_\text{rest} < 1180$~\AA{}. The quasar sightlines are chopped and padded by white noise based on the noise vector of the spectrum before passing into the VP-fitting program to avoid any complications arising from the edges of the spectra,
and the padded regions are later masked in post-processing. Such a treatment to the edges is also applied to the mock forward models to ensure our analysis is consistent.

\begin{table*}
\centering
\caption{Summary of HST STIS observations used in the study}
\begin{tabular}{ccccccc}
\hline
ID  &  $z_{qso}$   & STIS wavelength range & Obs date & Exp time & average SNR/pix& average SNR/pix\\ 
  & &(\AA{}) & &  (ksec) & (full spectra) &(\lya{} regions)\\
\hline
TON153      & 1.014 & 2275 - 3110 &  2001 Jan.  & 5.3  & 5.0 & 4.8\\
           & &                    &  2002 Jun.  & 8.2   \\
\hline                                  
PG1248+401  & 1.033 & 2275 - 3110 &    2002 Jul. &25.2  &  5.9 & 5.0\\
 & & & 2001 Oct.& 28.8 & & \\
\hline
PG1718+481  & 1.083 & 1841 - 2673 & 1999 Nov. & 14.1  & 7.9  & 9.8\\
\hline
PG1206+459$^{a}$  & 1.162 & 2273 - 3110 & 2001 Jan. & 17.3 & 7.3 & 6.4  \\
\hline
LBQS1435-0134$^{a}$ & 1.309 & 1985 - 2781 & 2015 Jun. & 20.9 & 10.6  & 5.5 \\
\hline
PG1241+176  & 1.283 & 2275 - 3110 &   2002 Jun.&  19.2 & 4.7 &  4.4 \\
\hline
PG1522+101$^{a}$  & 1.328 & 1985 - 2781 &  2015 Mar. & 7.7  & 9.5  & 7.1\\
            &      &   & 2015 May.  & 13.2 &  &\\
\hline
PG1634+706  & 1.337 & 1858 - 2673 &  1999 May. &  14.5 & 12.9  &18.7\\
            &      &  2275 - 3110 &  1999 Jun. &  14.5 & &\\
            &      &  1858 - 2673 &  1999 Jun. &  26.4 & &\\
\hline
PHL1377$^{a}$ & 1.440 & 2275 - 3110 &  2002 Jan.& 14.0 & 7.2 & 5.3 \\
 & &  & 2002 Feb.& 28.0 & &\\
\hline
PG1630+377$^{a}$  & 1.476 & 2275 - 3110 &    2001 Feb. &5.3   & 10.6 &  7.5\\
 & & & 2001 Oct.& 28.8 & & \\
\hline
PG0117+213  & 1.493 & 2275 - 3110 &   2000 Dec.&  42.0 & 7.2 & 7.5 \\
\hline
HE0515-4414 & 1.713 & 2275 - 3110 &  2000 Jan.  & 31.5  & 7.9  & 7.6\\
\hline
\end{tabular}
\raggedright
\begin{tablenotes}
      \small
      \item $^{a}$ The quasar sightlines on which we use the metal identification from the COS Absorption Survey of Baryon Harbors (CASBaH). 
    \end{tablenotes}
\label{tab_spec}
\end{table*}

\subsection{Voigt-Profile Fitting}
\label{sec:vpfit}

In this work, we use the line-fitting program \vpfit{}, 
which fits a collection of Voigt profiles convolved with the instrument LSF to spectroscopic data \citep[][]{vpfit}\footnote{VPFIT: \url{http://www.ast.cam.ac.uk/~rfc/vpfit.html}}. 
We employ a fully automated \vpfit{} python wrapper adapted from \citet{Hiss2018},
which is built on the \vpfit{} version 11.1. The wrapper routine controls \vpfit{} with the help of the \vpfit{} front-end/back-end programs \texttt{RDGEN} and \texttt{AUTOVPIN} and fit our simulated spectra automatically.
We set up \vpfit{} to explore the range of parameters $1  \leq  b  \leq 300~\text{km/s} $ and $11.5 \leq \log ( N_{\mathHI} / \text{cm}^{-2}) \leq 18$ for every single \lya{} absorption lines. 
\vpfit{} automatically varies these parameters and fits for additional component lines until the $\chi^2$ with respect to the whole spectral segment is minimized. 
Such a VP-fitting procedure is applied to the whole spectral segment, fitting both the \lya{} lines and metal lines, 
including both intervening metal lines and those from interstellar medium of \ac{MW}; for simplicity, 
hereafter we refer to these collectively as metal lines.
The removal of these metal lines is later discussed in \S~\ref{sec:metal_ID}.

During our VP-fitting procedure, we noticed the presence of artefacts in the spectra,
which are absent in our simulated and forward-modelled mock datasets.
A visual assessment of these minor features in the data suggested they were not genuine, but rather artefacts from factors like flat fielding, continuum placement, or data reduction artifacts. 
This is especially true for top-quality spectra, where the exceptionally high \ac{SNR} technically requires the inclusion of such faint components.
To this end, we incorporate a fixed 'floor' of 0.02 in quadrature to the normalized flux error vector for all spectra. We do this without introducing extra noise to the normalized flux. This value was determined through a process of trial and error. 
It was informed by the detection of a considerable number of absorption lines with notably low Doppler parameters and column densities, as reported by \vpfit{} in the spectra with the highest \ac{SNR}. 
These faint, narrow lines were not observed in our simulated and forward-modelled sightlines. 
Nevertheless, implementing this noise floor primarily affects lines with $\log \NHIt{}/\text{cm}^{-2}<12.5$ from our dataset, which will not be used for inference. This same noise floor is also applied to the simulated datasets to keep our data processing procedure consistent (see \S~\ref{sec:FM}).

 \begin{figure*}
\centering
    \includegraphics[width=0.95\linewidth]{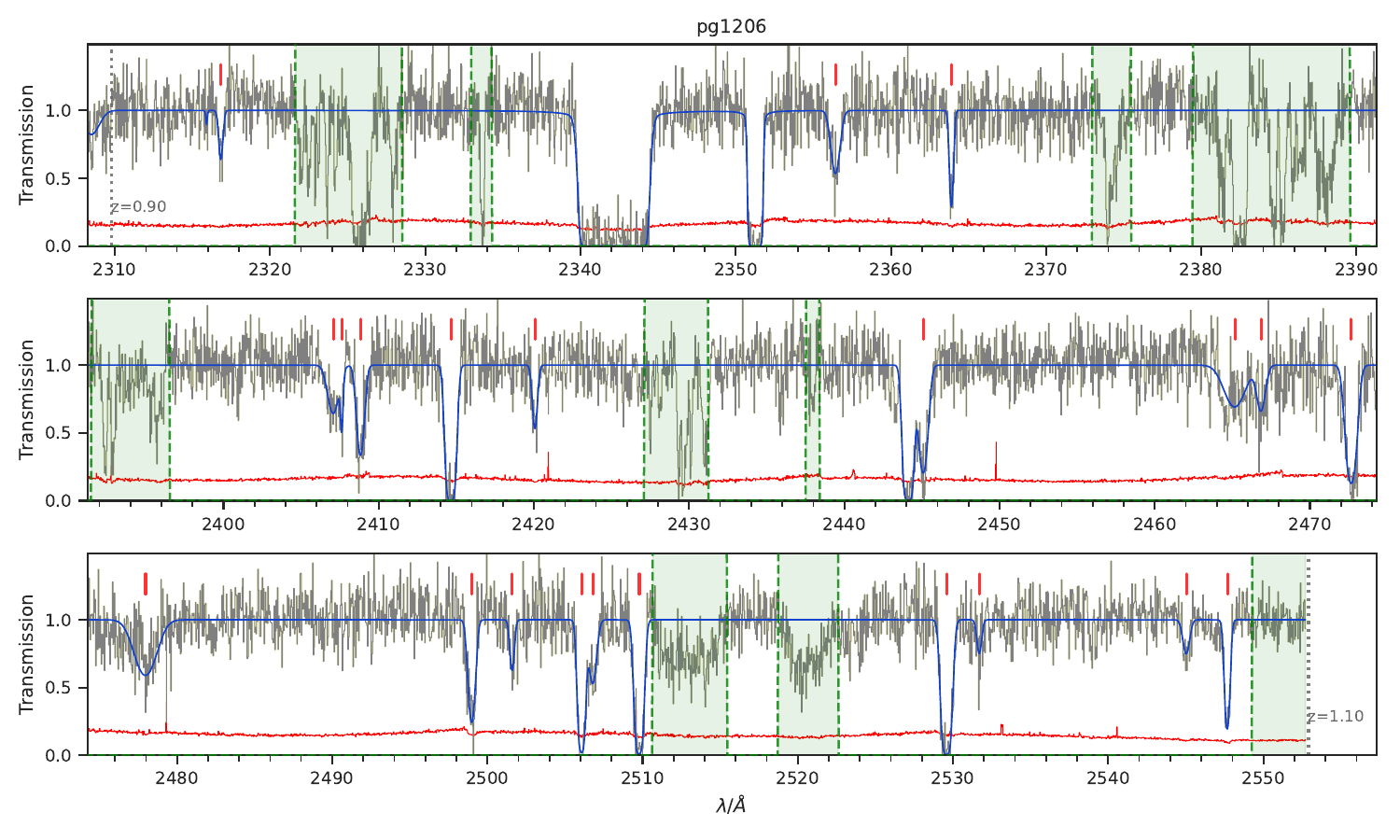}
  \caption{Illustration of the processed STIS spectrum of PG1206+459. 
  The original spectrum is shown in gray, while a model spectrum based on VP-fitting is shown in blue. 
  The noise vector is shown in red, and the masked regions are shown as green shaded regions. 
  The \lya{} lines used for our \bn{} dataset are labelled by red vertical lines.}
  \label{fig:spec_vp}
\end{figure*} 

Our \vpfit{} wrapper is designed to fit spectra using a custom \ac{LSF}.
However, it is important to note that it accommodates only a single \ac{LSF}, without accounting for any wavelength dependency. 
To address this, we extract the STIS E230M \ac{LSF} from {\tt linetools} and interpolate it to match the central wavelength of the spectrum we aim to fit. As previously detailed in \S~\ref{sec:data}, the STIS 230M exhibits a Gaussian-like \ac{LSF}, which shows minimal variation across different wavelengths. Consequently, our approach of employing a singular LSF in the VP-fitting process does not introduce significant errors. 
To ensure consistency and avoid statistical biases, we apply the same fitting methodology to both our observational data and forward-modelled mock.

Furthermore, we follow the convention used in previous studies \citep{schaye2000,rudie2012,Hiss2018}
and apply another filter for both  $b$ and $ N_{ \mathHI}$ in this study,
using only $b$-$ N_{ \mathHI}$ pairs in region $12.5 \leq \log ( N_{\mathHI} / \text{cm}^{-2}) \leq 14.5$ and $0.5 \leq \log ( b / \text{km s}^{-1}) \leq 2.5$ in our analysis.
Such a limitation is chosen to include the \bndist{}s for all our Nyx simulation models (see \S~\ref{sec:Thermal_para})
while guaranteeing that the absorption lines are not strongly saturated,
which maximizes the sensitivity to \ac{IGM} thermal state and minimizes the impact of poorly understood strong absorption lines arising mainly from the circumgalactic medium of intervening galaxies.

One of our STIS spectra, PG1206 is shown as an example of the VP-fitting procedure in Fig.~\ref{fig:spec_vp}. The original spectrum is shown in grey, and the model based on VP-fitting is shown in blue. The noise vector of the original spectrum is shown in red, and the masked regions due to metal line detection are shown as green shaded regions. 
The \lya{} lines used for our \bn{} dataset (after all filters) are labelled by red vertical lines.

\subsection{Metal Identification}
\label{sec:metal_ID}

As previously mentioned, our VP-fitting procedure fits all absorption lines including \lya{} lines and metal lines.
For our analysis based on the \bn{} of the \lyaf{}, 
it is critical to filter out these metal lines.
To this end, we make use of archival metal identification data presented in \citet{Milutinovic2007} for seven of our quasar sightlines 
and use metal identification from the CASBaH survey
\citep{CASBaH_data,CASBaH_OVI,CASBaH_NeVIII,CASBaH2021} for the rest five spectra (see notes of the Table.~\ref{tab_spec}).
For each spectrum, we create a mask to cover the vicinity of each metal line based on the aforementioned metal identification. 
These masked regions are initially aligned with the central wavelength of the metal lines reported in the literature,
while their initial widths are set to be $\Delta v=30$ km/s in velocity space.
Such a value is chosen based on the resolution of STIS E230M, which corresponds to $\sim 10$ km/s.
We then apply the masks to our VP-fit results to filter out potential metal lines.
To do so, we first locate the absorption line region characterized by $F_\text{line,fit} \leq$ 0.99,
where the $F_\text{line,fit}$ stands for the normalized flux given by the VP-fit model (the blue line in Fig.~\ref{fig:spec_vp}). 
If any absorption line region overlaps with the initial mask,  
we increase the width of the mask to cover the detected line,
while the increment is given by the full width at half maximum (FWHM) of the detected line, 
approximated by FWHM = $b$/0.6, where the $b$ is given by \vpfit{}.
Lastly, we adjust the masks manually to fill the small gaps (with $\Delta v=30$ km/s) between the masked regions and make sure
all absorption lines close to (the original) metal masks
reported by our VP-fitting procedure are masked.
The aforementioned masking procedure is needed based on the fact that our VP-fitting procedure does not match the line identified in the literature exactly,
due to the different spectra\footnote{ HST COS spectra are used in \ac{CASBaH} project to identify the metal lines.} used for metal identification and different post-processing procedures,
including coaddition, continuum fitting, and data smoothing used in our data. 
The aforementioned masking procedure makes sure that all potential metal contamination is removed. 
Afterwards, we manually masked a few gap regions in our quasar spectra resulting in the failure of the \vpfit{} caused by Damped \lya{} absorption systems (DLAs). 
These masks are generated in post-processing, which means that we first apply \vpfit{} to the spectra assuming all lines are \HI{} \lya{} and remove the absorption lines that fall within the masked regions, same as done for finding overlapped lines with metal masks. 
In the end, we subtract the metal mask from our total pathlength and obtain $\Delta z=$2.097.
Our full sample of quasar segments and their corresponding masks are presented in Appendix~\ref{sec:data_masks}. 

With our imposed cuts on the \bn{}, 
we find that 40 out of 341 lines are masked for our whole sample, 
and that leaves us with a \bn{} dataset consisting of 301 \lya{} absorption lines. 
We divide the 301 \lya{} absorbers into three redshift bins: $0.9 < z <1.1$, $1.1 < z < 1.3$ and $1.3 < z <1.5$ centered at $z=1$, $1.2$ and $1.4$, respectively, according to their central wavelength as determined by \vpfit{}. 
This provides us with the number of \lya{} lines to be 102, 160 and 39 and redshift path of 0.762, 0.972 and 0.363 in the bins centred at  $z=1$, $1.2$ and $1.4$, respectively.
In Table~\ref{tab_abs} we summarize our \bn{} dataset for each redshift bin, with 
redshift pathlength, number of final \lya{} lines as well as 
median values for the $b$ and $N_{\mathHI}$ in each bin.
\begin{table}
\caption{Summary of the of the observational dataset}
\centering
\renewcommand{\arraystretch}{2}
\begin{tabular}{ccccc}
\hline
$z$ bins  & $\Delta z$ & Number   & $b_\text{m}/ \text{km s}^{-1} $ & $\log (N_{{\mathHI},\text{m}} /  \text{cm}^{-2})$ \\ 
\hline
$0.9 \leq z \leq 1.1 $ &0.762  & 102 &31.74 & 13.48\\
$1.1 < z \leq 1.3 $ &0.972  & 160 &28.83 & 13.37\\
$1.3 < z \leq 1.5 $ &0.363  & 39 &29.69 & 13.48\\
\hline
\end{tabular}
\raggedright
\begin{tablenotes}
      \small
      \item The numbers of identified \lya{} lines in each redshift, the total pathlength $\Delta z$, 
and the median value $b_\text{m}$ and $\log N_{{\mathHI},\text{m}}$. 
    \end{tablenotes}
\label{tab_abs}
\end{table}

\section{Simulations}
  \label{sec:simulations}
  
 \begin{figure*}
 \centering
\includegraphics[width=1.0\textwidth]{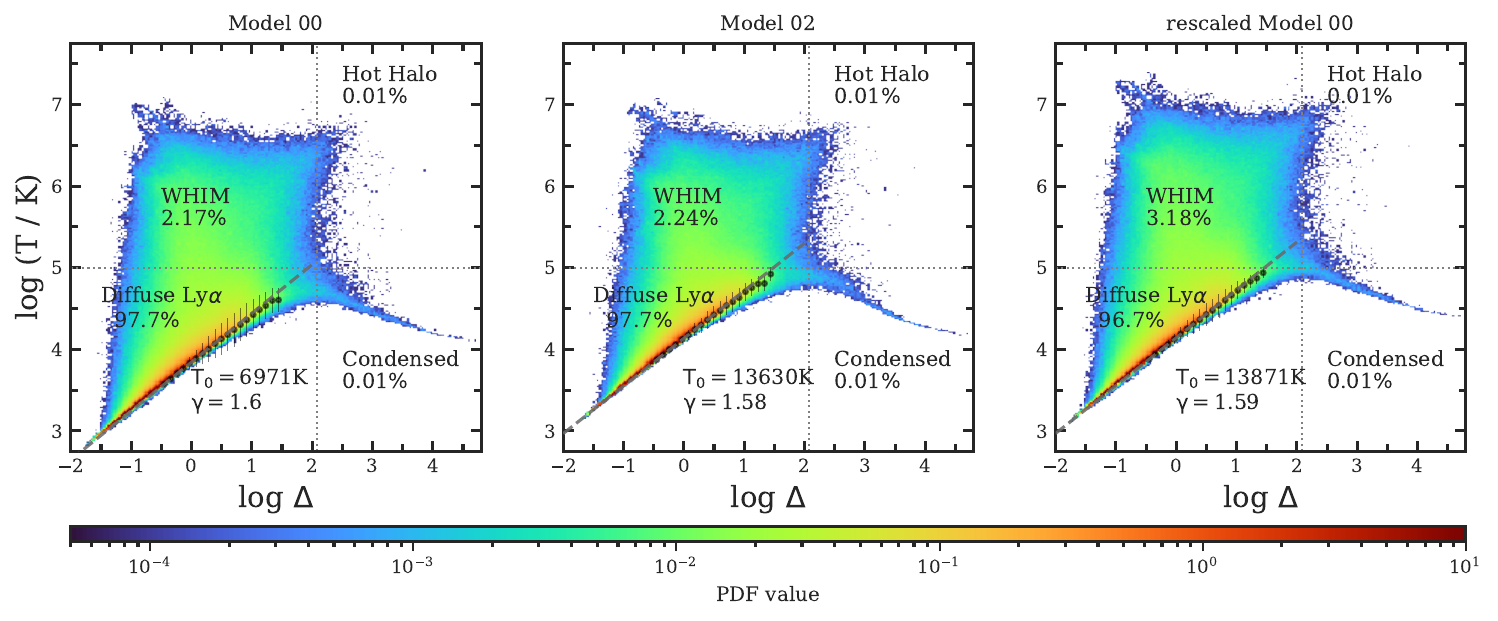}
  \caption{Volume weighted $T$-$\Delta$ distribution for three simulations models at $z=1.0$. 
The left panel is the Nyx model 00 with $T_0=6971$ K, $\gamma=1.60$, 
and the middle panel is the Nyx model 02 with $T_0=13630$ K, $\gamma=1.58$.
The right panel shows the model generated by multiplying the temperature in model 00 by two,
resulting in a $T_0 = 13871$ K and $\gamma =$1.59 according to our $\Delta-T$ fitting procedure.
The best-fit power-law relationship is shown as grey dashed lines. 
The $\log T$ for each bin are plotted as black dots, with the 1-$\sigma_{T}$ error bars shown as black bars. 
The volume-weighted gas phase fractions are shown in the annotations. 
The fraction of diffuse \lya~gas and the values of $T_0$ and $\gamma$ in the rescaled model (the right panel) agree within a percent level to the actual model shown in the middle panel. }
  \label{fig:Nyx_Rho_T}
\end{figure*} 

We utilize a set of Nyx cosmological hydrodynamic simulations \citep[see][]{Lukic2015, Almgren2013} to model the low-redshift \ac{IGM}. Developed primarily for simulating the \ac{IGM}, Nyx is a massively parallel cosmological simulation code. Within Nyx, dark matter evolution is captured by treating it as self-gravitating Lagrangian particles. In contrast, baryons are represented as an ideal gas on a uniform Cartesian grid, modelled using an Eulerian approach. The Eulerian gas dynamics equations are addressed using a second-order piece-wise parabolic method, ensuring accurate shock wave representation.

Nyx includes the main physical processes relevant for modelling the \lyaf{}. 
Nyx assumes the gas to have a primordial composition: a hydrogen mass fraction of 0.76, a helium mass fraction of 0.24, and zero metallicity. The various processes, such as recombination, collisional ionization, dielectric recombination, and cooling, are implemented according to the methodologies described in \citet{Lukic2015}. 
Nyx also models the process of inverse Compton cooling against the cosmic microwave background, 
and tracks the total thermal energy loss resulting from atomic collisional processes.
%
The default model of NyX uses spatially uniform UVB form \citet[][]{H&M2012}.
In subsequent stages, while generating the \lya{} forest in post-processing (See \S\ref{sec:skewers}), the UVB is treated as a variable parameter. Notably, since the Nyx simulations are tailored to study the IGM, they do not incorporate feedback or galaxy formation processes. This omission considerably reduces computational demands, enabling us to execute a vast ensemble of simulations with varied thermal parameters (as detailed in \ref{sec:Thermal_para}).

In this study, each Nyx simulation starts at $z=159$ and runs until $z=0.03$. It spans a simulation domain of $L_{\text{box}} = 20~{\rm cMpc}\slash h$ having $N_{\text{cell}} = 1024^3$ Eulerian cells for baryon and equal count of dark matter particles. The chosen box size strikes a balance between managing computational resources and ensuring convergence to within a margin of $< 10\%$ on small scales (reflected in large $k$ values). 
A more detailed discussion on resolution and box size considerations can be found in \citet{Lukic2015}.

Due to the inherent degeneracy between the thermal and ionization states of the IGM, it is essential to employ a substantial collection of Nyx simulations, each representing varied thermal histories. We provide a detailed description of the simulation grid utilized in this study in the following section.

 \subsection{Thermal Parameters and Simulation Grid}
   \label{sec:Thermal_para}

\subsubsection{The THERMAL suite}
\label{sec:THERMAL}
To represent the \ac{IGM} with varied thermal state spanning $0.9 < z< 1.5$, we utilize a subset of the \ac{THERMAL}\footnote{Details of the THERMAL suite is given in http://thermal.joseonorbe.com.} suite of Nyx simulations \citep[also see][]{Hiss2018,WaltherM2019}. From this suite, we use 51 models, each showcasing different thermal histories. For every model, we produce three simulation snapshots at $z= 1.0, 1.2$, and 1.4. From these, we determine the thermal state, characterized by [$\log T_0$,$\gamma$]. Varied thermal histories are realized by manually adjusting the photoheating rates ($\epsilon$), in accordance with the methodology set forth in \cite{Becker2011}.
In this method, $\epsilon$ is treated as a function of overdensity, i.e. 
\begin{equation}
\epsilon = A \epsilon_{\rm HM12} \Delta^B, 
\label{eqn:heating}
\end{equation}
where $\epsilon_{\rm HM12}$ represents the photoheating rate per  H~{\sc ii} ion,
tabulated in \cite{H&M2012}, and $A$ and $B$ are parameters used to generate models with 
different thermal histories.

As the universe evolves towards lower redshifts, the thermal state of the IGM tends to stabilize, making it a challenge to produce models with uniformly distributed $T_0$ and $\gamma$ values. For an in-depth discussion on this, refer to \citealt{WaltherM2019}. Specifically, crafting models with a low $T_0 (< 10^{3.5}~{\rm K})$ paired with a high $\gamma$ value $(> 1.9)$ at lower redshifts proves particularly daunting. When lowering the $T_0$ by decreasing the photoheating rates, the cooling resulting from the Hubble expansion starts to play a dominant role, causing $\gamma$ to gravitate towards a value close to 1.6 \citep[as discussed in][]{McQuinn2016}.
Consequently, the grid representing the interplay between $T_0$ and $\gamma$ assumes an irregular shape,
leaving voids in regions characterized by high $\gamma$ and low $T_0$. 
This irregularity also stems from the inherent design of the parameter grid in the \ac{THERMAL} suite,
which can be traced back to the thermal state analysis of higher redshifts \citep[][]{WaltherM2019}.

\subsubsection{$T_0$-rescaling models}
\label{sec:rescale}
As will be discussed later in \S~\ref{sec:result}, 
our data favour models with high $T_0$ at $z=1.0$ and $z=1.4$,
which is hard to generate based on the aforementioned procedure. 
This is because, as suggested by Eq.~\ref{eqn:heating}, our method alters the IGM thermal history of the simulation model by varying the heat released by the  H~{\sc i} photoionization. 
However, the results of such a heating procedure fade away in low $z$,
where the IGM is dominated by the adiabatic cooling caused by Hubble expansion \citep{McQuinn2016review}.  
As a result, the $T_0$ of the IGM at $z < 1.5$ becomes insensitive to the heat input in our method for models with high $T_0$.
To this end, we rescale the IGM temperature to model the IGM with high temperature. 
For $z=1.0$, we select six simulation snapshots with 3.75 $ \leq \log T_0 \leq$ 3.9, 
which has $T_0$ close to the Nyx model 00 with $A=1, B=0$ (see Eq.~\ref{eqn:heating}) at $z=1.0$,
and multiply their temperature $T$ (at each simulation cell) by 2.5 and 3 respectively to generate 12 new models. 
The other properties of the simulation remain unchanged, and since we rescaled the temperature of all simulation cells uniformly the whole $\Delta$-$T$ distribution of the simulation model still follows the power law  $\Delta$-$T$ relationship Eq.~\ref{eqn:rho_T} with the $T_0$ rescaled.
The [$T_0$, $\gamma$] of original models and models with rescaled $T_0$ are illustrated in Fig.~\ref{fig:corner_z10}, 
where the original models are shown as green dots,
the model rescaled to $2.5 \times T_0$ and $3.0 \times T_0$ are shown in orange and red respectively. 
Such temperature rescaling procedures are also applied to $z=1.4$ models, where our preliminary results also favour hot models, and the corresponding models are shown in Fig.~\ref{fig:corner_z14}. 

\subsubsection{Measuring the IGM thermal state $[T_0, \gamma]$}
\label{sec:measure_rho_T}

To measure the thermal state for each of the 51 models, 
we fit temperature-density ($T$-$\Delta$) relation (see Eq.~\ref{eqn:rho_T}) to the
temperatures and densities in the simulation domain.
While fitting the $T$-$\Delta$ relationship, 
we noticed broader distributions of the \ac{IGM} temperatures in low redshift ($z \lesssim 1.0$) compared to high redshift ($z>3$).
To accommodate the dispersion in the IGM $T$-$\Delta$ distribution while fitting the power-law relationship, we adopt the fitting approach detailed in \citetalias{Hu2022}. This method first segregates the diffuse \lya{} gas \citep[ $T <10^5$K and $\Delta <120$, see][]{Dave2010} into 20 bins based on $\log \Delta$. A linear least squares fit is then applied to the average temperatures within each bin. For this study, we've adjusted the fitting range to $-0.5 < \log \Delta < 1.5$. 
Examples of the $\Delta-T$ distribution and the corresponding power law fitting are shown in Fig.~\ref{fig:Nyx_Rho_T}.
For each panel, the best-fit power-law relationship is shown as grey dashed lines, 
and the $\log T$ for each bin are plotted as black dots, and the 1-$\sigma_{T}$ error bars are shown as black bars.
The left panel shows the Nyx model 00 with $T_0=6971$ K, $\gamma=1.60$, 
and the middle panel shows the Nyx model 02 with $T_0=13630$ K, $\gamma=1.58$ generated by varying the parameter $A$ and $B$ in Eq.~\ref{eqn:heating}.
The right panel shows the rescaled model 00 generated by multiplying the temperature in model 00 by two.
It exhibit a $T_0 = 13871$ K and $\gamma =$1.59 according to our $\Delta-T$ fitting procedure.

\subsubsection{ Varying the UVB $\Gamma_{\mathHI{}}$}
\label{sec:Gamma_grid}
Since we want to measure the ionization state of the IGM, 
we let the \HI{} photoionization rate $\Gamma_{\mathHI{}}$ be a free parameter when generating \lyaf{} skewers from our simulations. 
As such, we add an additional parameter $\log \Gamma_{\mathHI{}}$ to our thermal grid,
extending it to [$\log T_0$, $\gamma$, $\log \Gamma_{\mathHI{}}$].
Such procedure is done in the post-processing of the simulation, at the time when the simulated slightlines are generated (see \S~\ref{sec:skewers}).
The value of $\Gamma_{\mathHI{}}$ we used in this study spans from
$\log (\Gamma_{\mathHI{}} /\text{s}^{-1})$ = -11.2 to -12.8
in logarithmic steps of $0.2$ dex, which gives 9 values in total. 
These values are fixed for all redshift bins.

\subsection{Skewers}
\label{sec:skewers}

In this study, we produce mock \lya{} spectra by calculating the \lya{} optical depth ($\tau$) along lines of sight, referred to as skewers for simplicity. Within each simulation model, we create a set of 15,000 skewers, aligned with the $x$, $y$, and $z$ axes of the simulation box, distributing 5,000 skewers per axis.
Properties necessary for the optical depth calculation are then extracted from each cell along these skewers. These properties include the temperature ($T$), overdensity ($\Delta$), and the line-of-sight velocity ($v_z$). Additionally, the hydrogen neutral fraction ($x_{\mathHI{}}$), crucial for synthesizing Ly$\alpha$ forest skewers, is determined by assuming ionization equilibrium. This calculation incorporates both collisional ionization, dictated by the gas temperature ($T$), and photoionization.
In our approach, $\Gamma_{\mathHI{}}$ is treated as a free parameter during the post-processing. Considering that Nyx does not simulate radiative transfer, we employ an approximation to model the self-shielding effect of the UV background in optically thick gas. Following the method outlined by \cite{Rahmati2013}, this involves attenuating $\Gamma_{\mathHI{}}$ in cells with dense gas to mimic the effects of self-shielding.

Utilizing the given values of $x_{\mathHI{}}$, $T$, $\Delta$, $v_z$, and $\Gamma_{\mathHI{}}$, we compute the optical depth $\tau$ in redshift space. This is achieved by summing the contributions from all cells in real space along the line-of-sight, employing the full Voigt profile approximation as outlined by \citet{Tepper-G2006}. 
The continuum normalized flux of the \lyaf{} along these skewers is then determined by $F= e^{- \tau}$. 
For each specified $\Gamma_{\mathHI{}}$, this entire procedure is repeated to recalculate the skewers.
Here, in our approach, we avoid the common practice of rescaling $\tau$ when generating skewers for different $\Gamma_{\mathHI{}}$ values, a method typically used at higher redshifts.
This is due to the differing nature of the IGM at low-$z$. 
Unlike the high-$z$ IGM, which is predominantly influenced by photoionization, the low-$z$ IGM contains a substantial proportion of shock-heated Warm-Hot Intergalactic Medium (WHIM) gas. It is thus necessary to recalculate the skewers so as to take the contribution from collisional ionized gas into account.

\subsection{Forward Modeling of Noise and Resolution}
\label{sec:FM}
 
In this paper, we aim to measure the thermal and ionization state of the IGM at $z \sim 1$.
To this end, we generate mock datasets with properties consistent with our STIS E230M quasar spectra,
which comprise 12 unique quasar spectra.

For low-$z$ \ac{IGM} with temperatures at mean density $T_0\sim 5000~{\rm K}$,
the $b$-values for pure thermal broadening (i.e. the narrowest lines in the Ly$\alpha$ forest) are $b \sim 9$~{\rm km/s}, corresponding to a full width at half maximum (FWHM)$ \sim b/0.6 \sim 15$ km/s. 
Such absorption features can not be fully resolved by \ac{STIS} which has a resolution of roughly 10 {\rm km/s}.
Thus, it is crucial to treat the instrumental effect carefully.
Therefore, we forward model noise and resolution to make our simulation results statistically comparable with the observation data. 
In practice, we make use of tabulated STIS E230M LSF obtained from  {\tt linetools} 
and noise vectors from our quasar sample.
For any individual quasar spectrum from the observation dataset,
we first stitch randomly selected simulated skewers without repetition to cover the same wavelength of the quasar and then rebin the skewers onto the wavelength grid of the observed spectra. 
Then we convolve the simulated spectra with the HST \ac{STIS} \ac{LSF} while taking into account the grating and slits used for that specific data spectrum. 
The \ac{STIS} \ac{LSF} is tabulated for up to 160 
pixels in each direction. We interpolate the \ac{LSF} onto the wavelengths of the mock spectrum (segment) to obtain a wavelength-dependent \ac{LSF}. 
Each output pixel is then modelled as a convolution between the input stitched skewers and the interpolated \ac{LSF} for the corresponding wavelength. 
Afterwards, the newly generated spectrum is interpolated to the wavelength of the selected \ac{STIS} spectra. The noise vector of the quasar spectrum is propagated to our simulated spectrum pixel-by-pixel by sampling from a Gaussian with $\sigma = \psi_i$, with $\psi_i$ being the data noise vector value at the i$^\text{th}$ pixel. In the end, a fixed floor of 0.02 in quadrature is added to the error vector for all simulated spectra to avoid artificial effects in post-processing, as discussed in \S\ref{sec:vpfit}.

\begin{figure*}
\centering
    \includegraphics[width=0.95\linewidth]{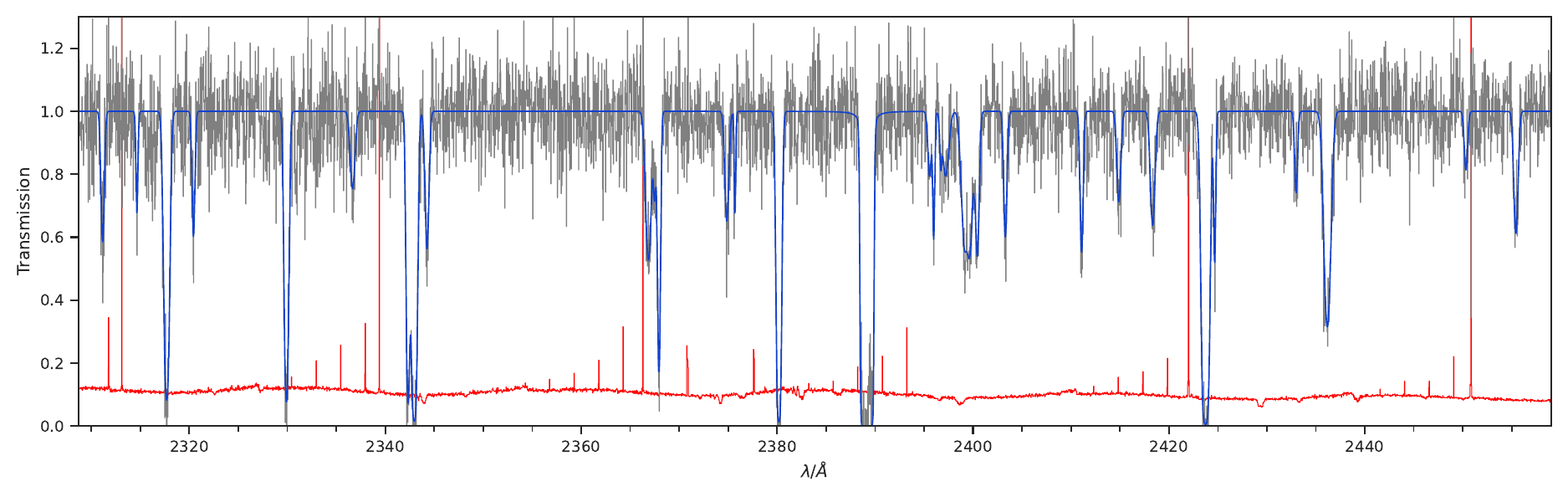}
  \caption{ An example of the mock spectrum (grey) forward-modelled from one of the STIS quasar spectra (segments), 
  with the corresponding noise vector (red). 
  The model fitted by \vpfit{} is shown in the blue. }
  \label{fig:model_VP}
\end{figure*}

For each model, including both Nyx model from the THERMAL suite and those generated by rescaling the temperature, we generated 1000 mock spectra, from the 15,000 raw skewers\footnote{
Generating 1000 spectra requires about 10,000 raw skewers, which are randomly selected from the total 15,000 skewers for each model.
}.
The total pathlength of the dataset for each model is roughly $\Delta z_\text{tot} \sim 100$.
We then fit Voigt profiles to each line in the spectra to obtain the \bn{} dataset used for the training of the \bndist{} emulator, which will be discussed in \S~\ref{sec:emu}.
For the purpose of illustration, an example of a forward-modelled spectrum is shown in Fig.~\ref{fig:model_VP} where the simulated spectrum is shown in grey, the model spectrum based on \vpfit{} line fitting is in blue, and the noise vector in red. 

 \section{Inference Method}
  \label{sec:inference}

 \subsection{Emulating the \bn{} Distribution}
 \label{sec:emu}
 
In this work, we make use of the inference framework following \citetalias{Hu2022},
which measures the thermal state and the photoionization rate $\Gamma_{\mathHI{}}$ of the low redshift \ac{IGM} using its \bndist{} and absorber line density \dndz{}.
The \bndist{} emulator is built on \ac{DELFI}, which turns inference into a density estimation task by learning the distribution of a dataset as a function of the labels or parameters \citep{papamakarios2016, Alsing2018, papamakarios2018, Lueckmann2018, Alsing2019}.
Following \citetalias{Hu2022}, we make use of \texttt{pydelfi}, the publicly available \texttt{python} implementation of \ac{DELFI},\footnote{See https://github.com/justinalsing/pydelfi} which makes use of \ac{NDE} to learn the sampling conditional probability distribution $P(\mathbf{d} \given \boldsymbol{\theta})$ of the data summaries $\mathbf{d}$, as a function of labels/parameters $\boldsymbol\theta$, from a training set of simulated data.
Here the data summaries $\mathbf{d}$ are [$\log \text{\NHI{}}$, $\log b$], and our set of label parameters $\boldsymbol{\theta}$ are the IGM thermal and ionization state [$\log T_0$, $\gamma$, $\log \Gamma_{\mathHI{}}$].

We generate training datasets by labelling the \bn{} pairs obtained from our 
mock spectra with the aforementioned labels.
We then train the neural network on the summary-parameter pairs $\{ [\log T_0,\gamma, \log \Gamma_{\mathHI{}}] , [b, \log \text{\NHI{}}]\}$.
Our \bndist{} emulator learns the conditional probability distribution $P ( b \mathbin{,} N_{\mathHI{}} \, \given \, T_0, \gamma, \log \Gamma_{\mathHI{}})$. These conditional \bndist{}s are then used in our inference algorithm, where we try to find the best-fit model given the observational/mock dataset, which is described in the following section. It is worth mentioning that we train our \bndist{} emulator for each redshift bin separately based on the corresponding training datasets.

\subsubsection{Likelihood function}
  \label{3sec:log-likelihood}

In Bayesian inference, a likelihood $\mathcal{L}= P(\mathrm{data}|\mathrm{model})$ is used to describe the probability of observing the data for any given model.
We adopt the likelihood formalism introduced in \citetalias{Hu2022},
which is summarized as follows,
\begin{equation}
\ln \mathcal{L} = \sum_{i=1}^{n} \ln (\mu_i) - \left(\frac{\text{d} N}{\text{d} z}\right)_{\rm model}\Delta z _{\rm data}, 
\label{eq:likelihood}
\end{equation}
where $\mu_i$ is the Poisson rate of an absorber occupying a cell in the $b$-$N_{\rm HI}$ plane with area $\Delta { \text{N}_{\mathHI{},i}}\times \Delta b_i$, i.e.
\begin{equation}
\mu_{i}=\left(\frac{\text{d} N}{\text{d} z}\right)_{\rm model}\,P(b_i, N_{\mathHI{},i} \given \boldsymbol{\theta})\,\Delta { N_{\mathHI{}}}\, \Delta b\, \Delta z _{\rm data}. 
\label{eq:mu}
\end{equation}
The $P(b_i,  \text{N}_{\mathHI{},i} \given \boldsymbol{\theta})$ in the equation
is the probability distribution function at the point $(b_i, N_{\mathHI{},i})$
for any given model parameters $\boldsymbol{\theta}$
evaluated by the DELFI \bndist{} emulator.
The $\Delta z_{\rm data}$ is the total redshift pathlength covered by the quasar
spectra from which we obtain our \bn{} dataset, and $\left({\text{d} N}\slash{\text{d} z}\right)_{\rm model}$ is the absorber density which is evaluated for any given set of parameters using a Gaussian process emulator (based on \texttt{George}, see \citealt{Ambikasaran2016}), which is also trained on our training datasets obtained from the Nyx simulation suite. 
More information on the likelihood function and the DELFI emulator can be found in \citetalias{Hu2022}.

\section{Results}
\label{sec:result}
\begin{figure*}
 \centering
    \includegraphics[width=.80\textwidth]{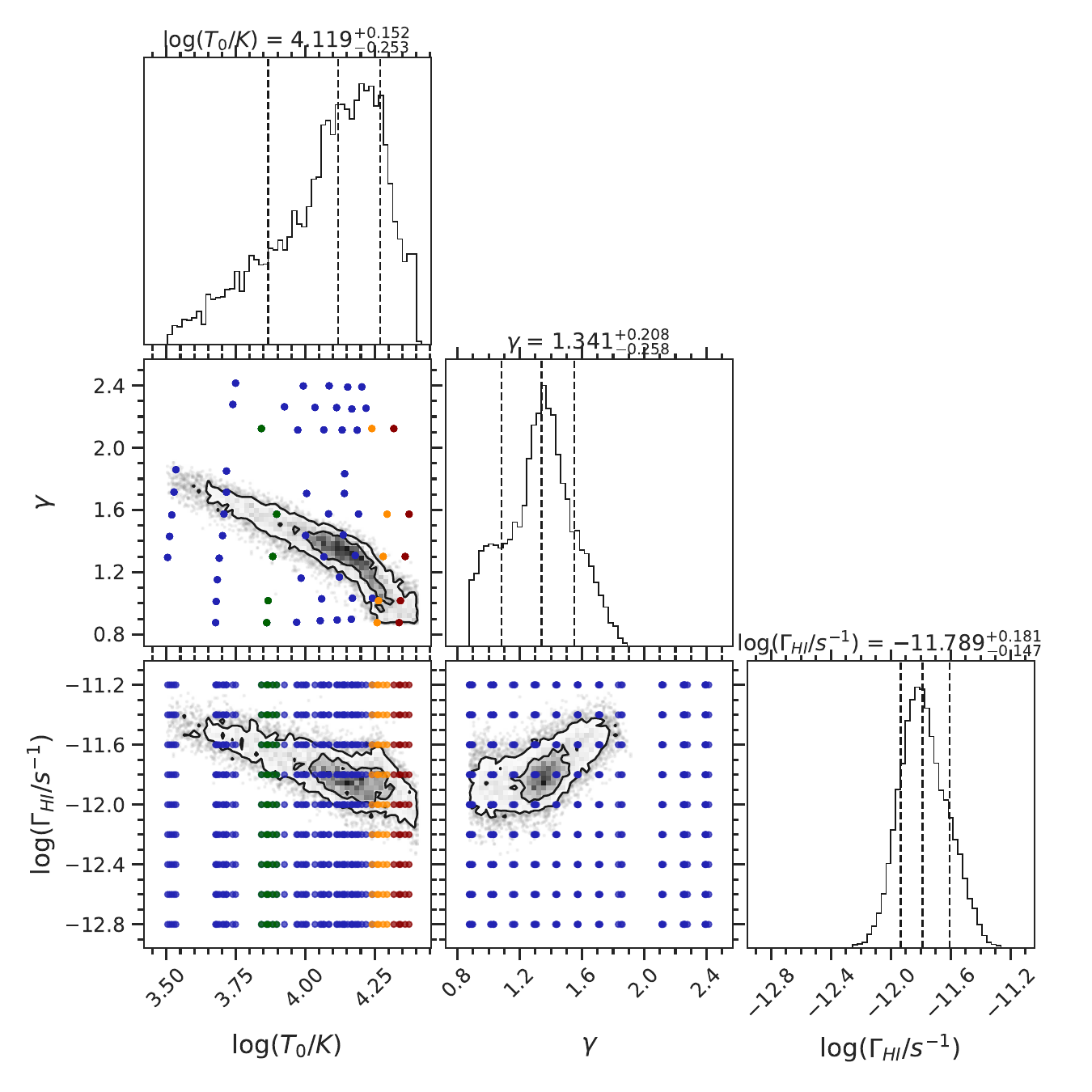}
  \caption{ The MCMC posterior obtained by our inference method using our \bn{} dataset at $z=1.4$. Projections of the thermal grid used for generating models are shown as blue dots.
  The Nyx models used for temperature rescaling are shown as green dots, and the models with 2.5 and 3.0 times $T_0$ are shown as orange and red dots respectively.
  The inner (outer) black contour represents the projected 2D 1(2)-sigma interval. The dashed black lines indicate the 16, 50, and 84 percentile values of the marginalized 1D posterior. }
  \label{fig:corner_z14}
\end{figure*}

\begin{figure*}
 \centering
    \includegraphics[width=.80\textwidth]{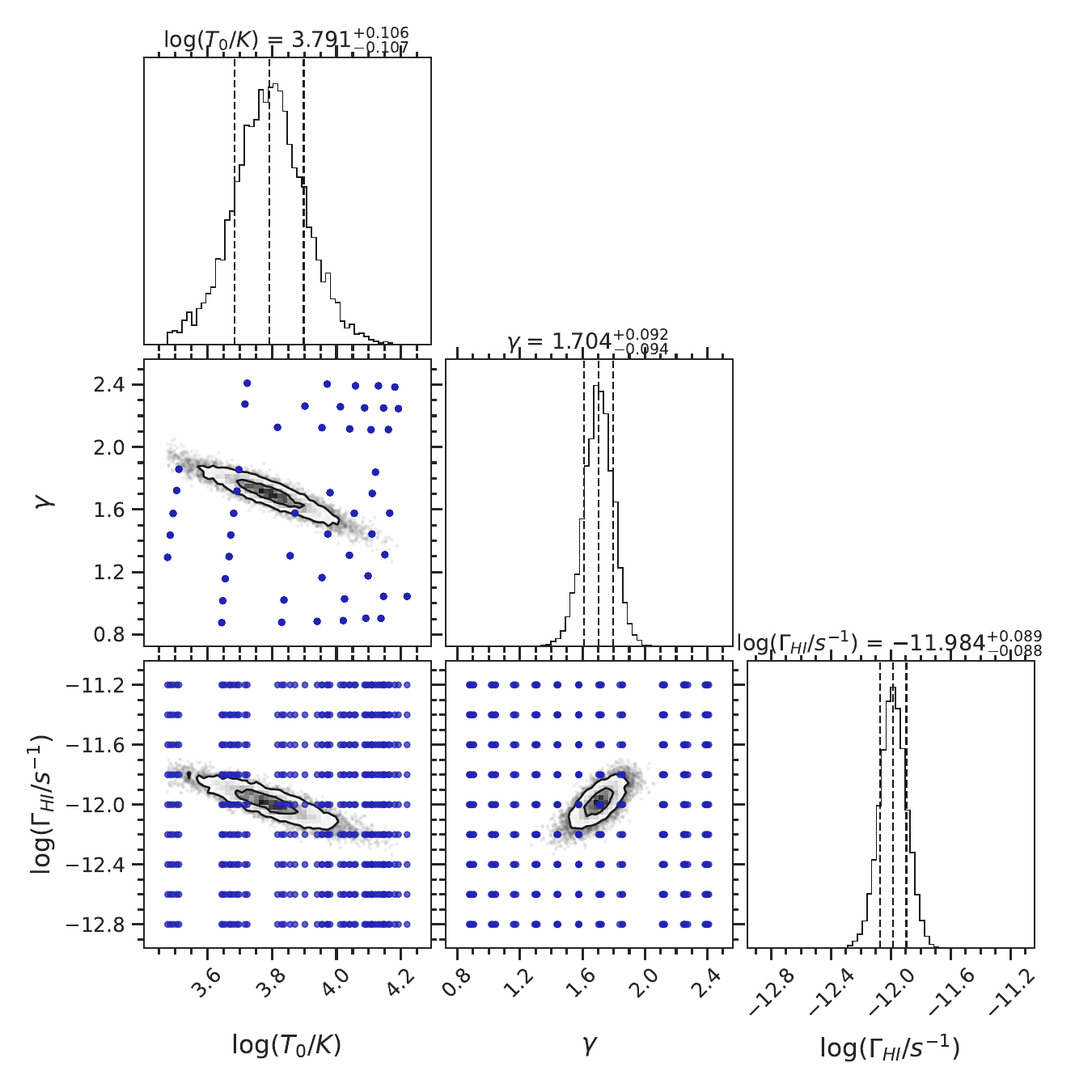}
  \caption{ The MCMC posterior obtained by our inference method using our \bn{} dataset at $z=1.2$. Projections of the thermal grid used for generating models are shown as blue dots, while the true model is shown as red dots. The inner (outer) black contour represents the projected 2D 1(2)-sigma interval. Red lines in the marginal distributions indicate the parameters of true models, while the dashed black lines indicate the 16, 50, and 84 percentile values of the marginalized 1D posterior. }
  \label{fig:corner_z12}
\end{figure*}

\begin{figure*}
 \centering

    \includegraphics[width=.80\textwidth]{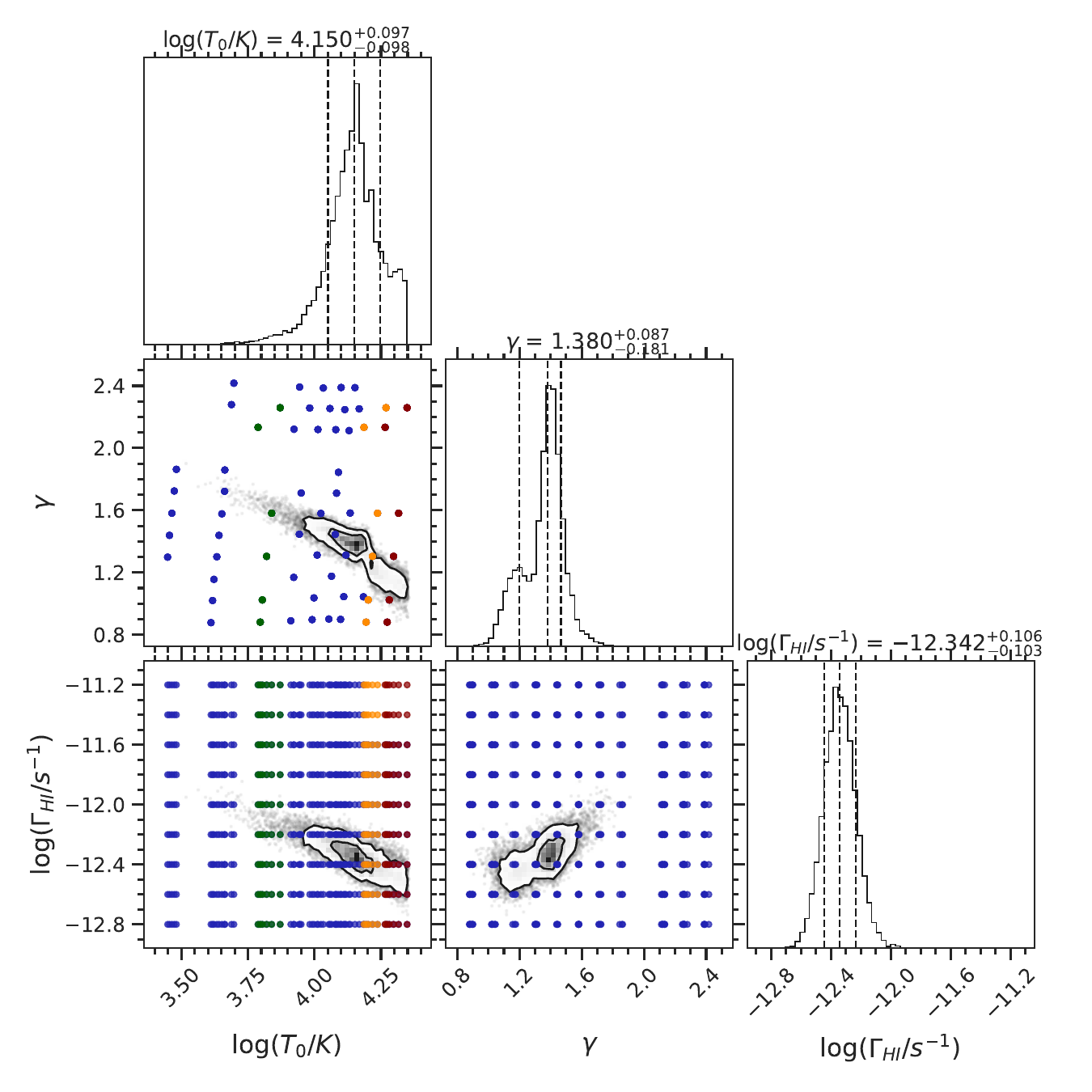}
  \caption{ The MCMC posterior obtained by our inference method using our \bn{} dataset at $z=1.0$. Projections of the thermal grid used for generating models are shown as blue dots. The Nyx models used for temperature rescaling are shown as green dots, and the models with 2.5 and 3.0 times $T_0$ are shown as orange and red dots respectively.
  The inner (outer) black contour represents the projected 2D 1(2)-sigma interval. The dashed black lines indicate the 16, 50, and 84 percentile values of the marginalized 1D posterior. }
  \label{fig:corner_z10}
\end{figure*}

We applied the aforementioned inference method to our dataset at three redshift bins to measure the IGM thermal and ionization state at $z=$ 1.4, 1.2, and 1.0. The resulting MCMC posteriors are presented in Fig.~\ref{fig:corner_z14}, Fig.~\ref{fig:corner_z12} and Fig.~\ref{fig:corner_z10} respectively. Projections of the thermal grid used for generating models are shown as blue dots. The inner (outer) black contour represents the projected 2D 1(2)-sigma interval. The dashed black lines indicate the 16, 50, and 84 percentile values of the marginalized 1D posterior. 
For $z=1.0$ and 1.4, our preliminary results indicate that the observational data favour models with high temperature, and the MCMC posterior is truncated at the boundary of the parameter space. 
As described in \S~\ref{sec:rescale}, these models with high temperatures are hard to model due to the heating mechanism used in the Nyx simulation. We thus manually rescale the temperature of some of the Nyx models and extend the parameter grid for our inference procedure as described in \S~\ref{sec:Thermal_para}.  
With these rescaled models, we are able to measure the thermal and ionization state of the IGM.
The parameter grids that contains the rescaled models are shown in Fig.~\ref{fig:corner_z14} and Fig.~\ref{fig:corner_z10} for z=1.4 and z=1.0 respectively.
The Nyx models used for temperature rescaling are shown as green dots, and the models with 2.5 and 3.0 times $T_0$ are shown as orange and red dots respectively.

\begin{figure*}
 \centering
     \includegraphics[width=1.0\textwidth]{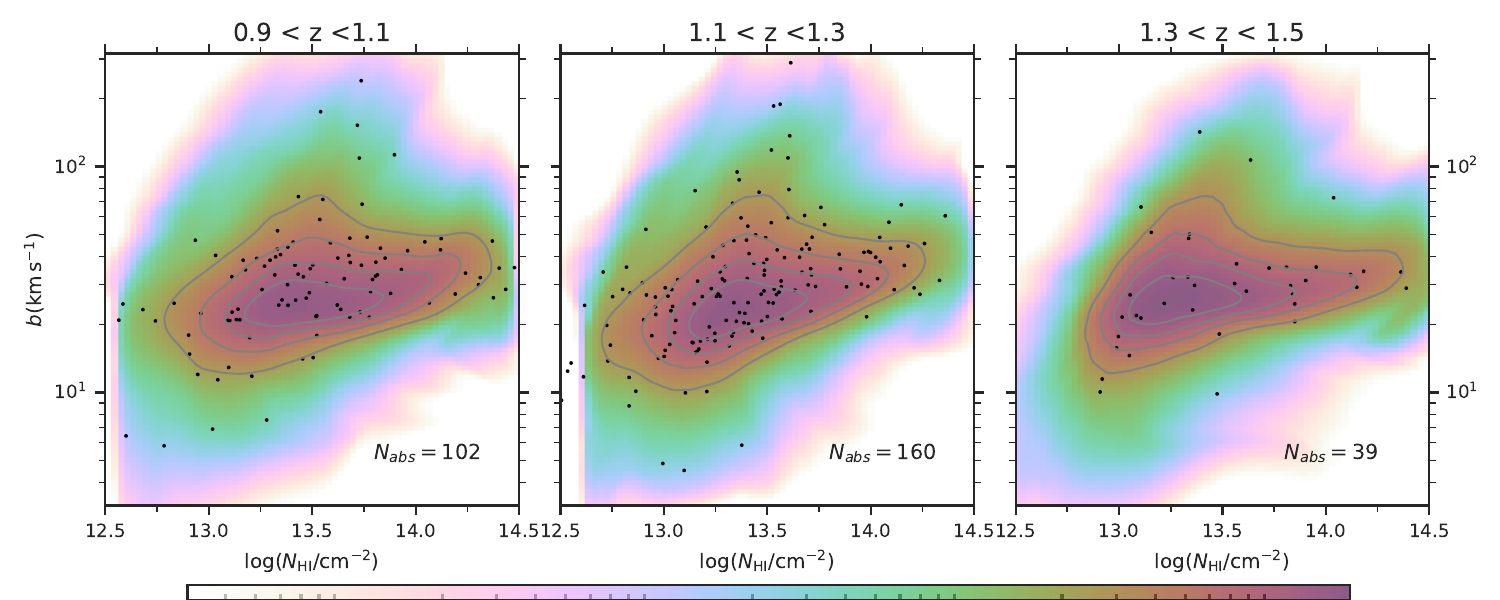}
  \caption{  Joint \bndist{}s emulated by our DELFI emulator based on the median values of the marginalized MCMC posterior at $z=$ 1.0, 1.2, and 1.4. Black dots are the \bn{} data. The likelihood contours corresponding to 80,60,40, and 20 cumulative percentiles CDF are plotted as gray solid lines. For illustration purposes, the values of the pdf are multiplied by 100 in the colour bar. }
  \label{fig:fit_evol}
\end{figure*}

We summarize the inference results (median values of the marginalized 1D posteriors for each parameter) in Table.~\ref{tab_inf_result}. 
\begin{table}
\centering
\caption{ Summary of the inference results}
\renewcommand{\arraystretch}{2}
\begin{tabular}{cccc}
\hline
$z$ bins &  $\log(T_0/\text{K})$  & $\gamma$ & log ($\Gamma_{\mathHI} / {\rm s}^{-1}$) \\ 
\hline
$1.3 < z \leq 1.5 $ & ${4.119}^{+0.152}_{-0.253}$ &${1.341}^{+0.208}_{-0.258}$ &${-11.789}^{+0.181}_{-0.147}$\\
$1.1 < z \leq 1.3 $ & ${3.791}^{+0.106}_{-0.107}$ &${1.704}^{+0.092}_{-0.094}$ &${-11.984}^{+0.089}_{-0.088}$ \\
$0.9 \leq z \leq 1.1 $ & ${4.132}^{+0.115}_{-0.103}$ &${1.357}^{+0.102}_{-0.151}$ &${-12.320}^{+0.103}_{-0.115}$ \\
\hline
\end{tabular}
\begin{tablenotes}
      \small
      \item The inference results i.e., median values of the marginalized 1D posteriors for each parameter, for all three redshift bins. The errors are given by the 1-$\sigma$ error (16-84\%) of the marginalized 1D posteriors.
    \end{tablenotes}
\label{tab_inf_result}
\end{table}
The \bn{} data and the corresponding \bndist{}s emulated by our DELFI emulator are shown in Fig.~\ref{fig:fit_evol}, 
and the likelihood contours corresponding to 80, 60, 40, and 20 cumulative percentiles are plotted as grey dashed lines. 
These plots show good agreement between the observational data and the emulated \bndist{}s.
We notice the inference result at $z=1.4$ has huge uncertainty due to the lack of observational data.
However, the precision is still satisfactory, given the fact that our \bn{} sample at this redshift bin contains only 39 data points. Such a size is comparable with the one used in \citet{ricotti2000}, whereas the error bar is much smaller (see Fig.~\ref{fig:Thermal_evol}), which mainly because of our novel method using full \bndist{} \citep[see][for the relevant discussion]{Hiss2019}.

\begin{figure*}
 \centering
     \includegraphics[width=1.0\textwidth]{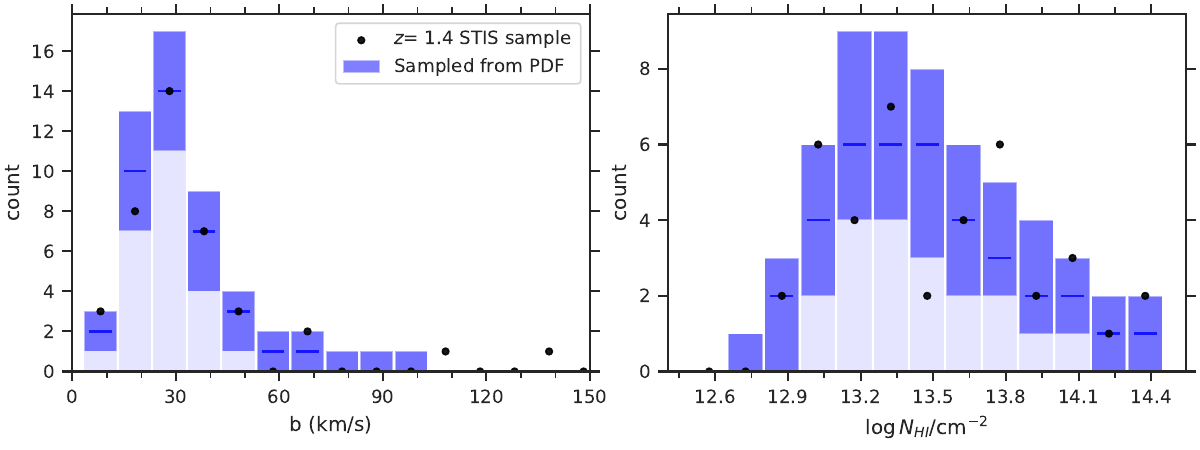}
  \caption{ The marginalized 1D $b$ and $N_{\mathHI}$ distributions of our are compared with 5000 mock datasets with the same size, sampled from the \bndist{}s emulated based on the median values of the MCMC posteriors. The black dots represent our \bn{} data at $z=$1.4. The blue bars indicate the mean value of the number of lines that fall in each bin for the 5000 datasets, whereas the blue shaded regions represent the 1-$\sigma$ uncertainty calculated from the 5000 datasets. }
  \label{fig:bN_1d_z14}
\end{figure*}

\begin{figure*}
 \centering
     \includegraphics[width=1.0\textwidth]{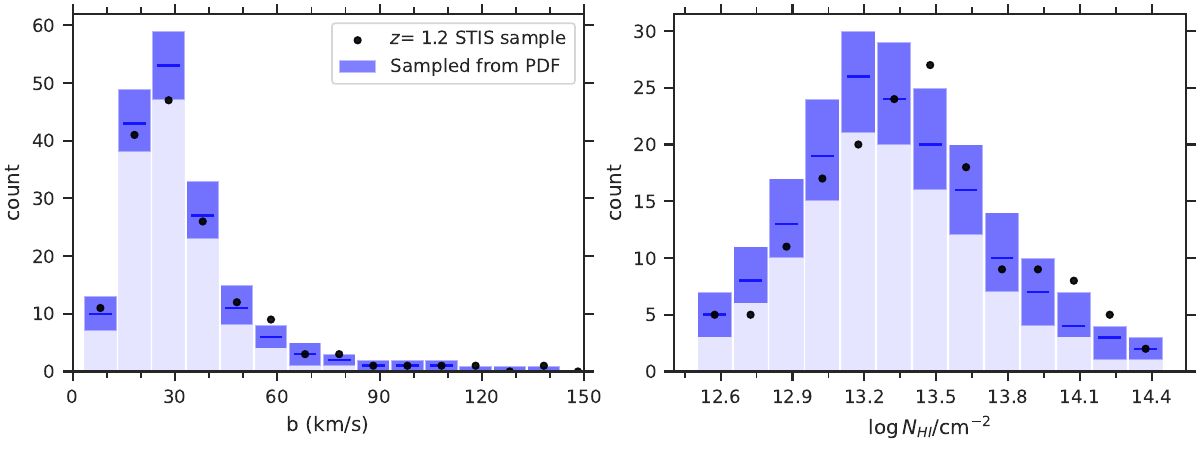}
  \caption{The marginalized 1D $b$ and $N_{\mathHI}$ distributions of our are compared with 5000 mock datasets with the same size, sampled from the \bndist{}s emulated based on the median values of the MCMC posteriors. The black dots represent our \bn{} data at $z=$1.2. The blue bars indicate the mean value of the number of lines that fall in each bin for the 5000 datasets, whereas the blue shaded regions represent the 1-$\sigma$ uncertainty calculated from the 5000 datasets.  }
  \label{fig:bN_1d_z12}
\end{figure*}

\begin{figure*}
 \centering
     \includegraphics[width=1.0\textwidth]{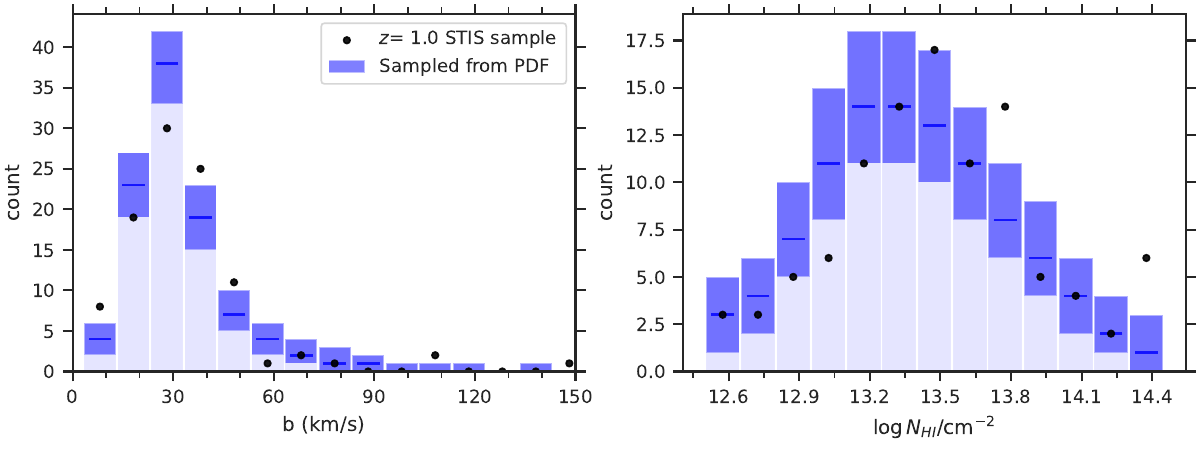}
  \caption{ The marginalized 1D $b$ and $N_{\mathHI}$ distributions of our are compared with 5000 mock datasets with the same size, sampled from the \bndist{}s emulated based on the median values of the MCMC posteriors. The black dots represent our \bn{} data at $z=$1.0. The blue bars indicate the mean value of the number of lines that fall in each bin for the 5000 datasets, whereas the blue shaded regions represent the 1-$\sigma$ uncertainty calculated from the 5000 datasets.  }
  \label{fig:bN_1d_z10}
\end{figure*}

Based on the marginalized 2D posteriors, we observe that our results across all redshift bins exhibit the anticipated degeneracies between parameters. Specifically, $T_0$ is degenerate with both $\gamma$ and $\Gamma_{\mathHI}$, as indicated in \citetalias{Hu2022}.
To further assess the goodness of our inference results, we plot the marginalized 1D $b$ and $N_{\mathHI}$ distributions of our sample in Fig.~\ref{fig:bN_1d_z14}, Fig.~\ref{fig:bN_1d_z12}, and Fig.~\ref{fig:bN_1d_z10} for each redshift bin, and compare them with 5000 mock datasets with the same size, sampled from the \bndist{}s emulated based on the median values of the MCMC posteriors. 
The blue bars indicate the mean value of the number of lines that fall in each bin for the 5000 datasets, whereas the blue shaded regions represent the 1-$\sigma$ uncertainty calculated from the 5000 datasets.
From the results, it is evident that our inference method adeptly recovers both the 2D and marginalized 1D distributions of \bn{}, even though the limited data size, particularly at $z=1.4$, leads to noticeable fluctuations, which are underscored by the substantial 1-$\sigma$ error bar in the marginalized 1D distributions in both $b$ and $N_{\mathHI}$ distributions.

As illustrated in Fig.~\ref{fig:bN_1d_z14}, Fig.~\ref{fig:bN_1d_z12}, and Fig.~\ref{fig:bN_1d_z10}, our 1D $b$-parameter distributions emulated for best fit [$T_0$, $\gamma$, $\Gamma_{\mathHI}$] are in good match with the observations, highlighting the robustness of our inference and suggesting that there is no severe discrepancy in $b$ distribution as opposed to the what is seen at $z<0.5$ \citep{Gaikwad2017, Viel2017}. 
Note that this $z<0.5$ $b$-parameter discrepancy arises from studies based on \ac{COS} low-$z$ \lya{} spectra \citep{Danforth2016}, 
however, in reality, the spectral resolution and LSF of COS may not be very good for accurate $b$-parameter measurements, especially for small $b$ values.
In contrast, old studies on higher-resolution STIS spectra, although with high uncertainty, 
found observed $b$-parameter in good agreement with predictions from cosmological simulations \citep[see Fig.~3 in][]{Dave&Tripp2001}{}. 
This consistency implies that the $b$-parameter discrepancy found in the literature may be an artifact of the limited spectral resolution provided by COS, which will be further investigated in our future work.
It also suggest that it might be beneficial to study the \lyaf{} with the higher-resolution spectra obtained with STIS. 

\subsection{Evolution of the thermal state of the IGM}
\label{sec:evol}

 \begin{figure*}
\centering
    \includegraphics[width=0.95\linewidth]{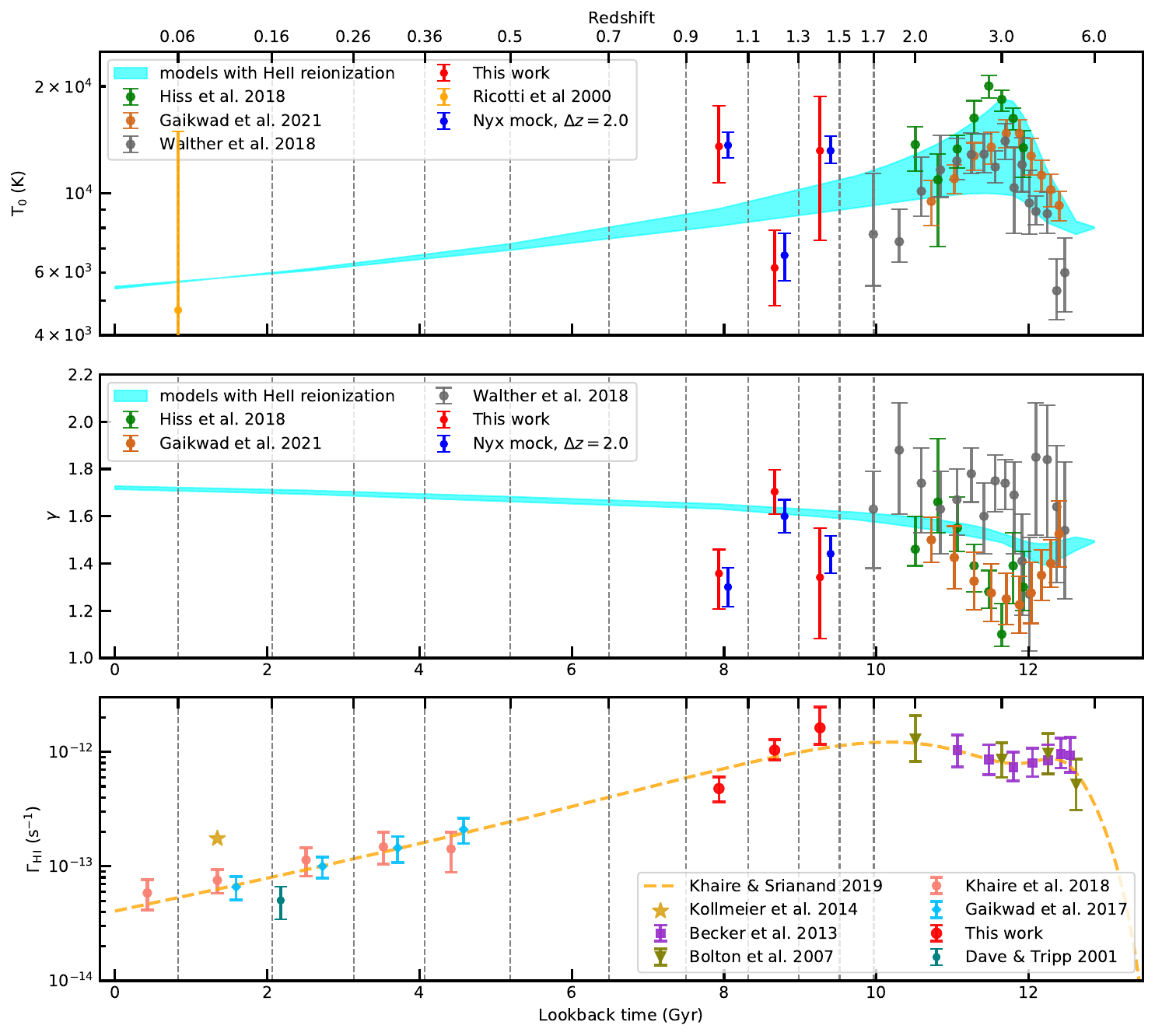}
  \caption{ Evolution history of $T_0$ (top) and $\log \Gamma_{\mathHI{}}$ (bottom) based on our inference results using the STIS data. Our results are shown as red dots, while measurements from other studies are displayed in different colours.
  The error bars stand for the 1-$\sigma$ error.
  The blue-shaded region in the top panel represents the range spanned by $T_0$ from hydrodynamical simulations of a large family of different HeII reionization models. The mock measurements based on Nyx simulation are shown in blue.}
  \label{fig:Thermal_evol}
\end{figure*} 

\begin{figure}
 \centering
    \includegraphics[width=0.5\textwidth]{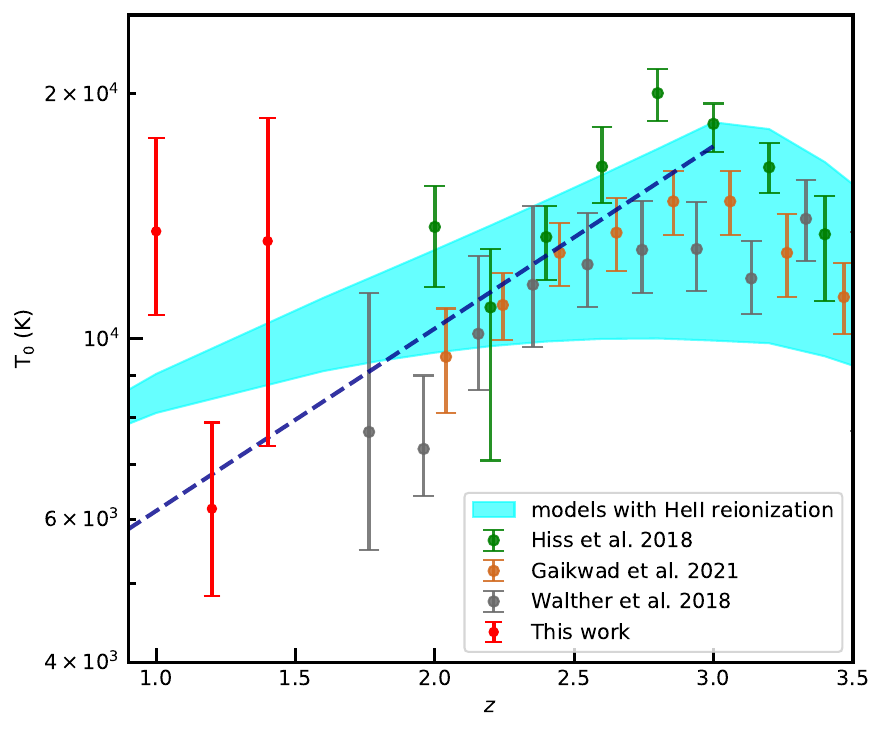}
  \caption{ The evolution of $T_0$ in the IGM across $0.9<z<3.5$, with results from previous studies shown in comparison. 
  The power law fit of $\log(T_0/\text{K})$ obtained by fitting all previous results in between $0.9<z<3.0$ are plotted as dark cyan dashed line.
  }
  \label{fig:T0_evol}
\end{figure}

In Fig.~\ref{fig:Thermal_evol} we summarize the $T_0$, $\gamma$ evolution across three redshift bins, 
and compare them with archival from previous studies at higher $z$
\citep{ricotti2000,Hiss2018,Walther2019,Gaikwad2021}.
Our results and their 1-$\sigma$ uncertainties are shown as filled red data points and error bars.
As a benchmark for current theoretical models,
we plot the IGM thermal history spanned by all potential Helium reionization models \citep{Onorbe2017,Onorbe2017b} as the cyan-shaded region.
To further assess how well do our low-$z$ results agree with previous results, 
in Fig.~\ref{fig:T0_evol},
we fit a power law relationship between $T_0$ and $z$ (blue dashed line), i.e., 
$ \log {T_0}(z) = c_1 z + c_2$, where $c_1, c_2$ are fitting coefficients obtained from a least squares linear fit based on all previous $T_0$ measurements in between $1.5 \leq z \leq3.0$ (i.e., not including our measurements).
Such a power law fit is a reasonable approximation in between $1.0 \leq z \leq 3.0$ \citep[see the prediction of low-$z$ $T_0$ in][]{Sanderbeck2016, McQuinn2016}.
The power law relationship (blue dashed line in Fig.~\ref{fig:T0_evol}) suggests that our measurements at $z=1.2$ and 1.4 are consistent with previous results. 
However, a noticeable discrepancy in $T_0$ emerges at $z=1.0$,
where our measurement of $T_0$~ 13500 K is significantly higher than best-fit power-law relationship predicted by previous measurements.
Such a discrepancy suggests that the IGM may be far hotter than expected at $z \sim 1.0$, 
implying the existence of extra heating sources that are not included in our current IGM model,
which becomes crucial at $z \sim 1.0$.  
Summarizing the $T_0$ measurements across all three redshift bins,
two potential thermal histories for the IGM emerge:
(1) The IGM might undergo a cooling phase around $z \sim 1.2$
before heating up to 13500 K at $z \sim 1.0$,
which is not unfeasible given the significantly large time span of $\sim 700$ Myr between these two redshifts. 
(2) Alternatively, the IGM could consistently maintain a high temperature since 
$z \sim 1.5$. 
However due to the substantial error bars in $T_0$ in all three redshift bins, 
no definitive conclusion can be made until further investigation with larger datasets.

\begin{figure}
 \centering
    \includegraphics[width=0.5\textwidth]{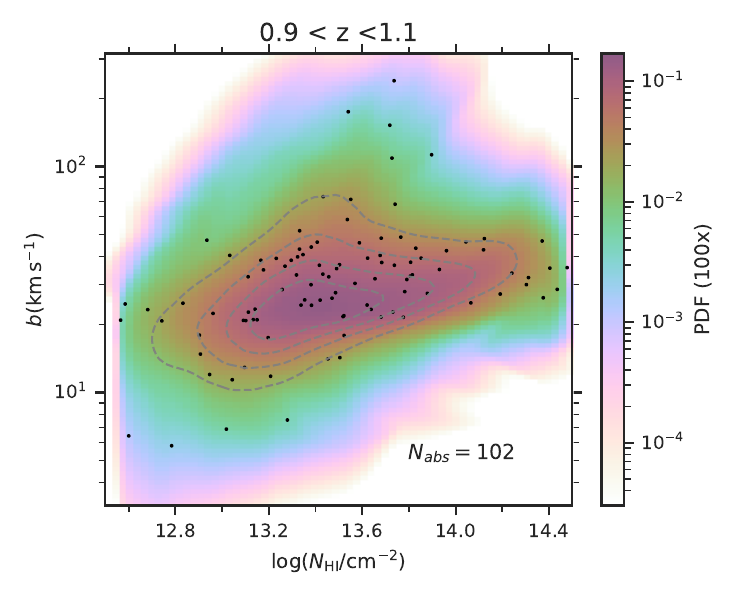}
  \caption{ The likelihood contours of the \bndist{} at $z=1.2$ on top of the \bn{} dataset and the corresponding \bndist{} at $z=1.0$.
  The likelihood contours corresponding to 80,60,40, and 20 cumulative percentiles are plotted as gray dashed lines. 
  }
  \label{fig:corner_compare}
\end{figure}

To further investigate the possibe change of the IGM thermal state from $z=1.2$ to $z=1.0$,
in Fig.~\ref{fig:corner_compare}, 
we over-plot the likelihood contours of the \bndist{} at $z=1.2$ on top of the \bn{} dataset and the corresponding \bndist{} at $z=1.0$.
It can be seen that the \bn{} dataset and the corresponding \bndist{} at $z=1.0$ lies above the likelihood contours of the \bndist{} at $z=1.2$,
suggesting that our observational data indeed favour a rapid change in the IGM thermal state between $1.0<z<1.2$.
More discussion on this unexpected high $T_0$ is present in \S~\ref{sec:T0_discrepancy}.

As for $\gamma$, our results for $z=1.4$ and 1.2 align with this trend
as outlined in \citet{McQuinn2016review},
in which the value of $\gamma$ tends to decrease towards lower redshifts.
However, the result at $z=1.0$ indicates a reduced $\gamma$.
The cause of this discrepancy remains unclear, 
but it is worth noting that such a trend of $T_0$ and $\gamma$, i.e., high $T_0$, low $\gamma$,
is consistent with the $T_0$-$\gamma$ degeneracy shown in the inference posterior
(see the 2D marginalized posterior contours in $T_0$-$\gamma$ plane in Fig.~\ref{fig:corner_z10}).
As a result, it is likely that the inference results at $z=1.0$,
which yields high $T_0$ and low $\gamma$ are caused by inference uncertainty and degeneracy.
On the other hand, 
it is also possible that the IGM starts to heat up at $z\sim 1.0$, 
leading to both increasing $T_0$ and decreasing $\gamma$.
In this case, the inconsistencies observed in both $\gamma$ and $T_0$ have a common root cause.

\begin{figure*}
 \centering
    \includegraphics[width=1.0\textwidth]{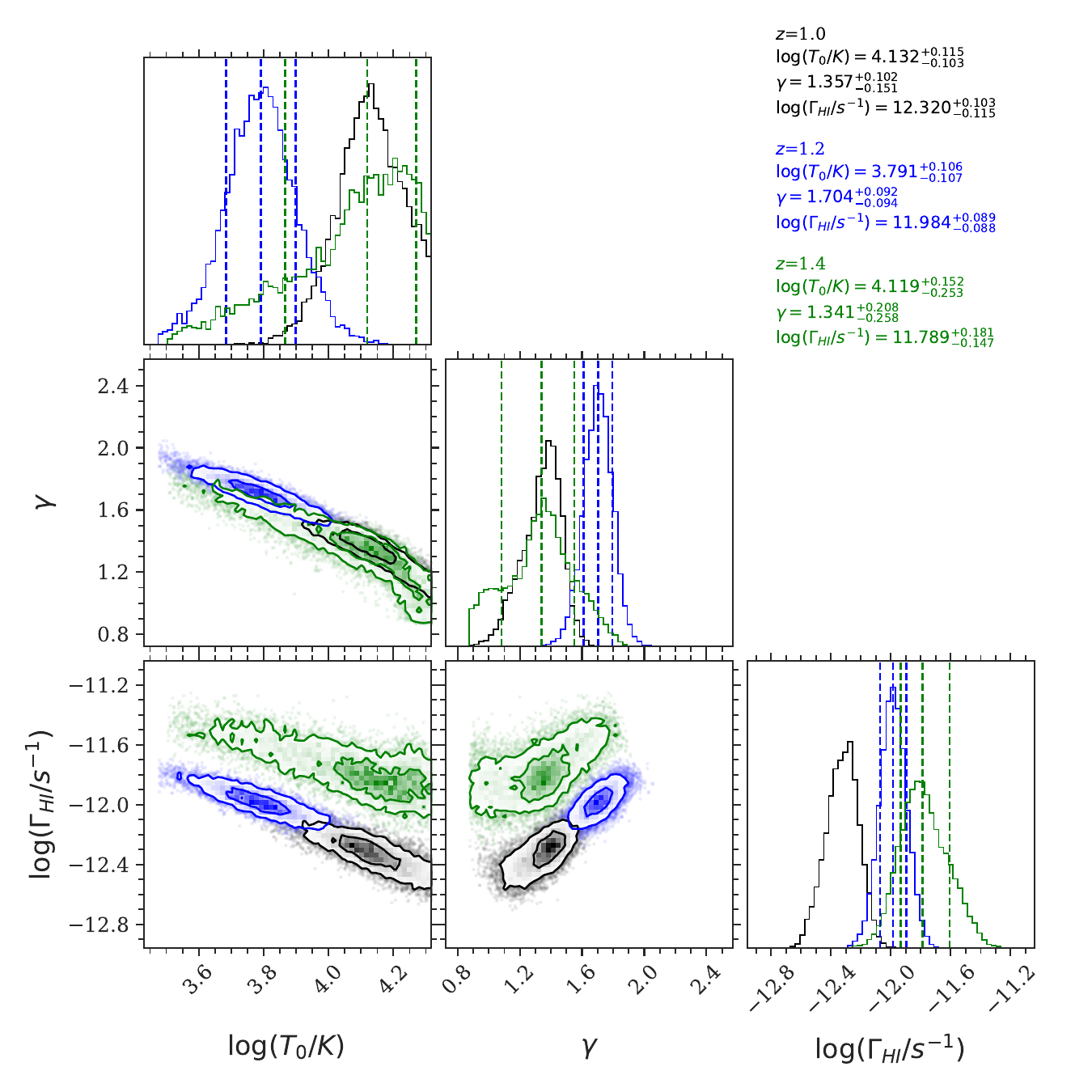}
  \caption{ The MCMC posteriors obtained by our inference method for all three redshift bins. The $z=1.0$ posterior is shown in black, the $z=1.2$ posterior is shown in blue, and the $z=1.4$ posterior is shown in green.}
  \label{fig:corner_evol}
\end{figure*} 

To illustrate the evolution of the IGM thermal and ionization state, 
we over-plot the three MCMC posterior on top of each other in Fig.~\ref{fig:corner_evol}, 
where the 2D marginalized posterior for $z=1.4$ are shown in green, 
the one for $z=1.2$ is plotted in blue, and the one for $z=1.0$ is shown in black. 
From the $T_0$-$\gamma$ plane, 
we observe a clear turnover for both $T_0$ and $\gamma$ at $z = 1.0$, 
suggesting a reverse evolution trend at $z \sim 1.0$. 
Such synchronization between the evolution of $T_0$ and $\gamma$ is important for us to understand the origin of the discrepancy,
and relevant discussion is presented in \S~\ref{sec:T0_discrepancy}.

\subsection{Evolution of the H~{\sc i} photoionization rate and the UVB}
\label{sec:evol_Gamma}

Our measurements fill in the $\Gamma_{\mathHI}$ evolution history between $0.0<z<1.7$.
In the bottom panel of Fig.~\ref{fig:Thermal_evol} we show our $\Gamma_{\mathHI}$ 
measurements across our three redshift bins, 
compared with previous studies
\citep{Dave&Tripp2001,Bolton2007,Becker2013,Kollmeier2014,Gaikwad2017, Khaire2019}.
Our inference results indicate that the $\Gamma_{\mathHI{}}$ is in good agreement with 
the UVB model presented in \citet{Khaire_Srianand2019} in all three redshift bins.  

It is worth
noting that for $z<3$, the UVB model of \citet{Khaire_Srianand2019} is dominated by 
photons emitted by quasars alone i.e., the escape fraction on ionizing photons from galaxies
is negligible at $z<3$. 
Our $\Gamma_{\mathHI{}}$ measurements support the same conclusion that galaxies are not the main source of ionizing photons at $z<3$. 
The same conclusion can be drawn from the new UVB models of \citet{Puchwein19} and  \citet{FG2020} because their $\Gamma_{\mathHI{}}$ values align very well with the UVB model of \citep{Khaire_Srianand2019} at $z<2$. This is mainly because all three UVB models use updated quasar luminosity functions at $z<3$ \citep[as presented in][]{Croom09, Ross13, PD13} after 
\citet{Khaire15puc} pointed out that previous UVB models \citep{H&M2012, Faucher-Gigu2009}
used old quasar luminosity function that predicts factor two smaller ionizing emissivity.

The consistency of our new $\Gamma_{\mathHI{}}$ measurements in the previously unexplored redshift range with recent UVB models attests to the robustness of these UVB synthesis models, especially in the aspect of hydrogen ionizing part of the UVB.

\section{Discussion}
\label{sec:discussion}

\subsection{The discrepancy in $T_0$}
\label{sec:T0_discrepancy}
In this section, we delve into the observed discrepancy in IGM thermal state at $z \sim 1.0$. 
First of all, we notice a coherence between the high $T_0$ measured at $z\sim 1$ and the high $b$-values observed at $z \sim 0.1$,
based on the \ac{COS} \lyaf{} dataset \citep{Danforth2016}, 
where the observed $b$-parameter significantly surpass the predicted value based on various simulations \citep{Gaikwad2017, Viel2017, Nasir2017, Bolton2021, Bolton22}.
Quantitatively, \citet{Viel2017} compares the marginalized $b$ distribution with various simulations, 
showing that the $b$ distribution at $z \sim$ 0.1 can be best recovered by the hydrodynamic simulations \citep[\texttt{P-GADGET-3}, see][]{Springel2005} with $ T_0 \gtrsim 10000$ K, 
while the theoretical model dictates that the $ T_0 \sim 5000$ at $z=0.1$. 
The similarity of required IGM temperature at both $z=0.1$ and 1.0
suggests that the discrepancy at $z \sim 0.1$ may be related the one at $z \sim 1$,
indicating a persistent trend from $z \sim 1.0$ to $0.1$.
Additionally, it also suggests that the discrepancy observed at $z=0.1$ may not be attributable to the limited resolution of the COS.

The simplest explanation for these discrepancies is the thermal broadening caused by a higher-than-expected IGM temperature,
which requires the existence of extra heating sources. 
If this is true,
our understanding of \ac{IGM} physics will be changed drastically,
highlighting a severe need to investigate processes that are possibly responsible for it,
such as dark matter annihilation \citep[][]{Araya2014, Bolton22}, 
gamma-ray sources \citep[][]{Puchwein2012},
or feedback from galaxy formation, 
whose effects are not fully understood in low-$z$ \citep[see][]{Springel2005,Croton2006,Sijacki2007,Hopkins2008,Tillman2023,Tillman2023b,Hu2023}.

Another possible explanation instead of extra heating is the presence of unexpected non-thermal broadening mechanisms affecting the $b$-parameter of the \lyaf{}, such as micro-turbulence motion in the IGM induced by jet or feedback \citep{Gaikwad2017, Viel2017, Nasir2017, Bolton2021}. 
However, these non-thermal broadening models fail to account for the unexpected trend in $\gamma$ observed in our results,
where the $\gamma$ are lower than expected at $z=1.0$.
To further investigate this, we plan to apply our inference method to the COS \lyaf{} dataset at $z \leq  0.5$, 
which should help to break the degeneracy between $T_0$ and $\gamma$,
thereby providing deeper insight into the $b$ discrepancy observed at $z \sim 0.1$.

\subsection{ Forecast based on mock observations}
\label{sec:forecast}

In this section, we make realistic forecasts for our future measurements with more abundant observational data. 
Given the amount of the newfound bright objects expected in upcoming surveys including \textit{Gaia} DR3 \citep{Gaia2016,gaia2023}. 
With a realistic amount of the observation from \ac{HST} \ac{STIS}, i.e., $\sim 50$ orbits,
we expect the path length coverage for each redshift bin to be significantly extended.
Here we assess the constraining power based on total pathlength $\Delta z=2$ for each redshift bin, 
corresponding to three times the current data size or roughly 15 spectra for each redshift bin,
while assuming the characteristic SNRs of the data do not change. 
We pick forward-modelled mock spectra from our mock dataset at each redshift bin and generate mock observational data with total pathlength $\Delta z$=2.
The Nyx model used here is the one with the thermal state that is closest to the inference results presented in \S~\ref{sec:result}.
The inference results obtained from these mock observations and their 1-$\sigma$ error bars are shown in Fig.~\ref{fig:Thermal_evol} as blue dots.
It can be seen that with $\Delta z = 2$, 
the 1-$\sigma$ errors for $T_0$ become roughly $\sim 1500$ K,
and the 1-$\sigma$ errors for $\gamma$ become roughly $\sim 0.08$.
These results will help us to confirm whether the IGM cools down as predicted.

\subsection{ The effect of potential contamination }
\label{sec:contamination}

In spite of the careful masking procedure, 
our \bn{} dataset still encounters potential contaminants, 
including blended lines and unidentified metal lines,
especially for the metal masks obtained from \citet{Milutinovic2007},
since their metal identification might not be complete. 
Here we briefly discuss the potential effects of these contaminants. 
It is well known that ionic metal line transitions mainly contribute to narrow absorption lines with $b \leq 10$ km/s \citep{schaye1999,Rudie2012a,Hiss2018}.
As a result, the metal line contaminants tend to bias our inference toward lower $T_0$. 
To this end, these contaminants shall not affect the main and most important result of this paper,
i.e., the IGM seems to be hotter than expected at low-$z$, especially at $z=1.0$.
For these blended lines, in this paper, we adopt a more conservative metal masking,
where we manually filter out all suspicious lines close to the masked regions 
(see the masks in Appendix~\ref{sec:data_masks}).
As for a more detailed quantitative analysis, 
we plan to identify all \lya{} lines using the \lyb{} (or higher transitions) forest \citep[see e.g. ][]{Rudie2012a}.
We plan to do this in future by combining our data set with other archival and upcoming data form HST.

\subsection{ The effect of SNRs of the spectra}
\label{sec:discuss_SNR}
We notice that a few quasar sightlines in our sample have relatively low SNRs (see Table.~\ref{tab_spec}),
and it is unknown whether our results are biased by these spectra.
Hence, in this section, we test the effect of these low-SNR sightlines on our inference results.
To do that, we exclude three quasar spectra from our sample 
which have relatively lower SNR ($\leq 5$), while the remaining spectra all have SNR $>$5.
We exclude TON153, PG1248+401 and PG1241+176 from the observational data and obtain a new \bn{} dataset,
which provides 25 fewer \lya{} lines compared with the old one and reduces the total pathlength $\Delta z$ by 0.24. 
We generate new mock datasets based on the nine spectra with SNR $>$ 7, 
and train our emulators based on the new dataset.
The outcomes indicate that even after excluding low SNR spectra from our data (and correspondingly in our mock data), 
the results remain consistent across each redshift bin. 
Such a result is important for our future work, 
suggesting that it is possible to make use of relatively low SNR data to obtain higher total pathlength and analyse the evolution of the thermal and ionization state on finer redshift bins, 
such that we could pinpoint the onset of the discrepancy in $T_0$ (or $b$-parameter) between the observation and simulation more precisely.

\section{Summary and Conclusion}
\label{sec:conclusion}

In this paper, we make use of 12 archival STIS E230M quasar spectra, 
from which we obtain the \bndist{} distribution and line density \dndz{} over the redshift range $0.9 < z < 1.5$ in three redshift bins.
We then measure the thermal and ionization state of the IGM following a
machine-learning-based inference method presented in \citetalias{Hu2022} for this redshift range for the first time.
Below we summarize our resutls:
\begin{itemize}
     \item We Voigt-profile fit the \lya{} in all 12 quasar spectra using a fully automated \vpfit{} wrapper and obtain \bn{} for 341 lines.
     We use the metal identifications from the \ac{CASBaH} project and combine them with the metal identification from \citet{Milutinovic2007} 
     to generate our metal masks,
     filtering out 40 contaminants besides \lya{} absorption lines, 
     and obtain a final sample of 301 \lya{} lines across a total pathlength of $\Delta z=$2.097. 

     \item We employ the \citetalias{Hu2022} inference method, 
     which simultaneously measures $[T_0, \gamma, \Gamma_{\mathHI}]$ from the \bndist{} and \dndz{}, 
     with the help of neural density estimators and Gaussian process emulators trained on a suite of 51 Nyx simulations each having a different IGM thermal history.
     It enables us to measure the IGM thermal and ionization state with high precision even with limited data.
    
     \item  We obtain $[\log T_0, \gamma] = [{4.119}^{+0.152}_{-0.253},{1.341}^{+0.208}_{-0.258}]$ at $z=1.4$ and $[\log T_0, \gamma]=[{3.791}^{+0.106}_{-0.107},{1.704}^{+0.092}_{-0.094}] $ at $z=1.2$.
     These two measurements agree with the theoretical model (Fig.~\ref{fig:Thermal_evol} and \ref{fig:T0_evol}), 
     suggesting that the thermal state of the IGM evolves as expected from $z=1.4$ to $z=1.2$.

     \item Nevertheless, our results yield $[\log T_0, \gamma]$ = [${4.132}^{+0.115}_{-0.103}$, ${1.357}^{+0.102}_{-0.151}$ ] at $z =1$, suggesting an unexpectedly high IGM temperature and low $\gamma$, 
     which is against the trend predicted by the current theoretical models of the IGM.
     Such high $T_0$ potentially suggests the existence of extra heating or unexpected non-thermal broadening at $z \sim 1.0$.

     \item Based on our measurements, it is possible that the IGM experiences a cooling phase until  $z \sim 1.2$ from $z\sim3$, and then it gets heated up to 13500 K at $z= 1$ in approximately $700$ Myr. Alternatively, the IGM temperature might have remained consistently high since $z \sim 2$.      
     However, due to significant uncertainties in $T_0$ for all three redshift bins,
     a definitive conclusion cannot be reached without further investigation.

     \item The inference results of $\gamma$ suggest that it also goes through unexpected evolution at $z \sim 1$.
     However, while it is likely that such a trend is caused by extra heating that causes the discrepancy in $T_0$,
     it is also possible that it is due to inference degeneracy between $T_0$ and $\gamma$.

     \item We compare our findings with previous work, which reports unanticipated high $b$-parameters compared with various simulations based on observational data at $z \sim 0.1$. These high $b$ values, if caused by thermal broadening, correspond to an IGM temperature with $T_0 \sim 10000$K. This convergence towards a higher IGM temperature aligns with our findings and suggests that the discrepancy in $b$-parameter observed at $z \sim 0.1$  \citep{Gaikwad2017,Viel2017} could be related to the one we have identified in this study. It further implies that the observed discrepancy may emerge around $z \sim 1.0$ and persist down to $z \sim 0$.
     
     \item We successfully measure the $\Gamma_{\mathHI}$ at three redshif bins,
     reporting $\Gamma_{\mathHI}={-11.789}^{+0.181}_{-0.147}$, ${-11.984}^{+0.089}_{-0.088}$, 
     and ${-12.320}^{+0.103}_{-0.115}$ at $z=1.4, 1.2$ and $1.0$ respectively.
     These measurements align well with the predictions of recent UVB synthesis models \citep{Khaire_Srianand2019, Puchwein19, FG2020}, reinstating the fact that low-z 
     UV background (at $z < 3$) is dominated by radiation from quasars alone.

     \item By excluding three spectra with relatively low SNRs from our quasar sample,
     we confirm that our results are not sensitive to the SNR of the dataset,
     suggesting that it is feasible to conduct our analysis on larger quasar samples with lower SNR to make finer measurements of the IGM thermal and ionization state, so as to pinpoint the onset of the discrepancy in the IGM thermal state between the observation and simulation more precisely. 

     \item We perform mock measurements using realistic datasets based on Nyx simulation to forecast the constraining power for our future work.
     The results demonstrate that with redshift pathlength 
     $\Delta z \sim 2.0$ for each redshift bin,
     three times the current data size,
     we will be able to constrain the $T_0$ within $\pm$ 1500 K.
     This precision will help us to constrain the thermal history of the IGM in $0.9<z<1.5$, and confirm whether the IGM cools down as expected at $z \sim 1.0$.

\end{itemize}

Alongside our quasar sample, we have acquired approximately 15 additional archival STIS spectra across $0.5 < z <2$,
which is roughly half the size needed to draw a comprehensive picture of the low-$z$ IGM thermal evolution.
After metal identification, these spectra can be incorporated into our analysis easily.
We are also in the process of obtaining more observational data from the HST.
Furthermore, we also planned to apply our inference methodology to simulations that utilize more sophisticated and diverse feedback mechanisms,  such as those featured in EAGLE \citep{EAGLE} and the CAMELS suite \citep{camels_presentation}. 
These efforts will deepen our understanding of how different feedback processes affect the low-$z$ IGM,
and help investigate the causes behind the discrepancies observed between simulations and observations in the thermal state of the low-$z$ IGM.

\begin{acronym}
	\acro{AGN}{active galactic nuclei}
	\acro{CMB}{Cosmic Microwave Background}
	\acro{COS}{Cosmic Origins Spectrograph}
	\acro{DELFI}{density-estimation likelihood-free inference}
	\acro{DM}{dark matter}
	\acro{DLA}{damped Ly$\alpha$}
	\acro{GP}{Gaussian process}
	\acro{HIRES}{High Resolution Echelle Spectrometer}
	\acro{HST}{Hubble Space Telescope}
	\acro{IGM}{intergalactic medium}
	\acro{KDE}{Kernel Density Estimation}
	\acro{KODIAQ}{Keck Observatory Database of Ionized Absorbers toward QSOs}
	\acro{LD}{least absolute deviation}
	\acro{LLS}{Lyman limit systems}
	\acro{LS}{least squares}
	\acro{LSF}{line spread function}
	\acro{MCMC}{Markov chain Monte Carlo}
	\acro{MW}{Milky Way}
	\acro{NDE}{neural density estimation}
	\acro{PCA}{principal component analysis}
	\acro{PDF}{probability density function}
	\acro{PKP}{\ac{PCA} decomposition of \ac{KDE} estimates of a \ac{PDF}}
	\acro{QSO}{quasi-stellar objects}
	\acro{SNR}{signal-to-noise ratio}
	\acro{STIS}{Space Telescope Imaging Spectrograph}
	\acro{TDR}{temperature-density relation}
	\acro{THERMAL}{Thermal History and Evolution in Reionization Models of Absorption Lines}
	\acro{UV}{ultraviolet}
	\acro{UVB}{ultraviolet background}
	\acro{UVES}{Ultraviolet and Visual Echelle Spectrograph}
	\acro{WHIM}{warm hot intergalactic medium}
        \acro{CASBaH}{COS Absorption Survey of Baryon Harbors}
\end{acronym}

\section*{Acknowledgements}

The authors thank the ENIGMA members\footnote{http://enigma.physics.ucsb.edu/} and Joe Burchett for helpful discussions and suggestions. 

Calculations presented in this paper used the hydra and draco clusters
of the Max Planck Computing and Data Facility (MPCDF, formerly known
as RZG). MPCDF is a competence center of the Max Planck Society
located in Garching (Germany).
This research also used resources of the National Energy Research Scientific Computing Center (NERSC), a U.S. Department of Energy Office of Science User Facility located at Lawrence Berkeley National Laboratory, operated under Contract No. DE-AC02-05CH11231 
In addition, we acknowledge PRACE for awarding us access to JUWELS hosted by JSC, Germany.
JO acknowledges support from grants BEAGAL18/00057 and CNS2022-135878 from the Spanish Ministerio de Ciencia y Tecnologia.

\section*{Data Availability}

The simulation data and analysis code underlying this article will be shared on reasonable request to the corresponding author. 



\bibliographystyle{mnras}


\bibliography{references.bib}


\appendix
\section{Observational data and Metal masks}
\label{sec:data_masks}

We present our observational spectra and the corresponding masks.
The original spectrum is shown in grey, and the model based on VP-fitting is shown in blue. The noise vector of the original spectrum is shown in red, and the masked regions due to metal line detection are shown as green shaded regions. The \lya{} lines used for our \bn{} dataset (after all filters) are labelled by red vertical lines.

 \begin{figure*}
\centering
    \includegraphics[width=0.95\linewidth]{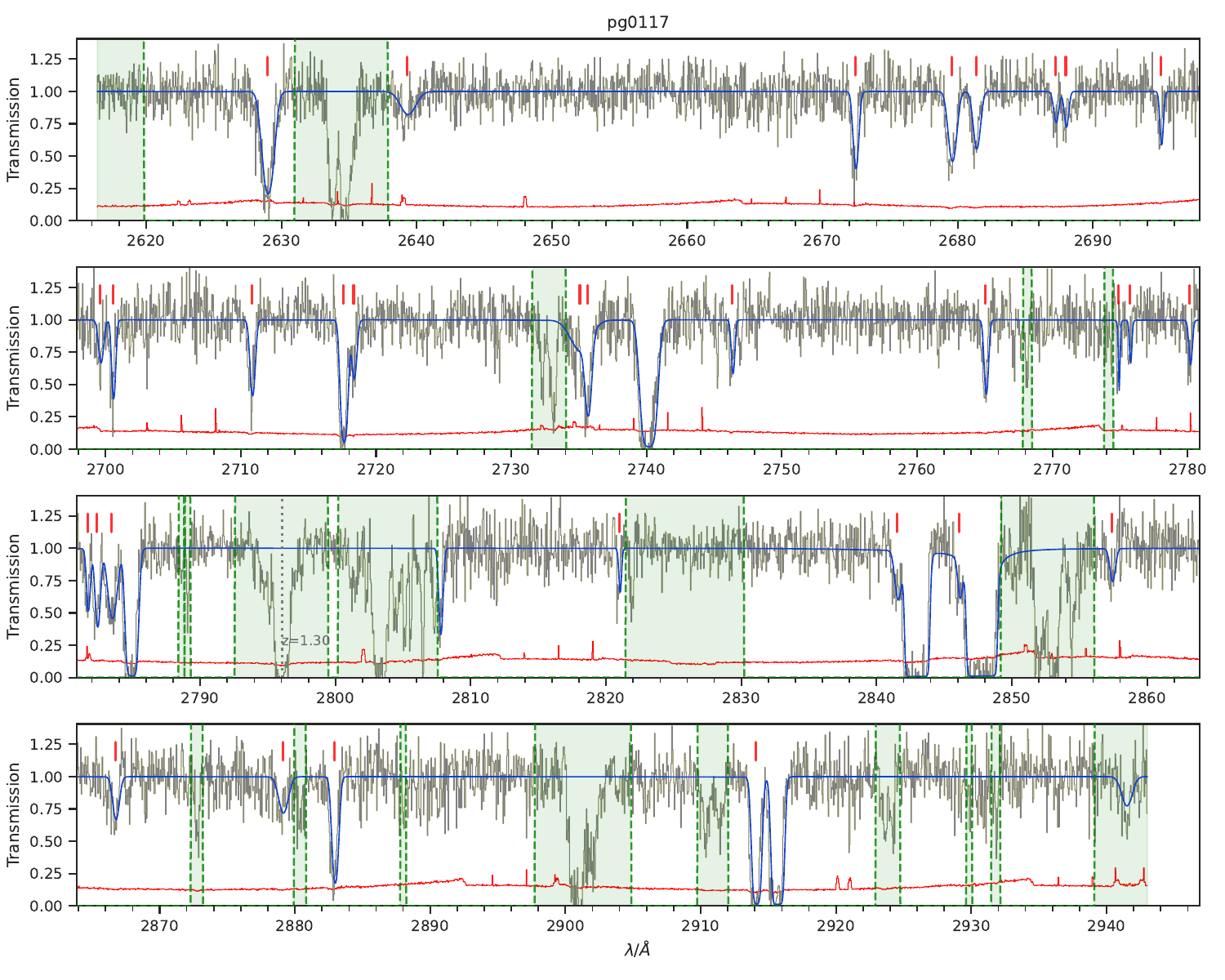}
  \caption{ 
  Illustration of the processed STIS spectrum of PG0117+213.
  The original spectrum is shown in gray, while a model spectrum based on line fitting (described in \S~\ref{sec:vpfit}) is shown in blue. The noise vector of the original spectrum is shown in red, and the masked regions are shown as a green shaded region. 
  The \lya{} lines used for our \bn{} dataset are labelled by red vertical line. }
     \includegraphics[width=0.95\linewidth]{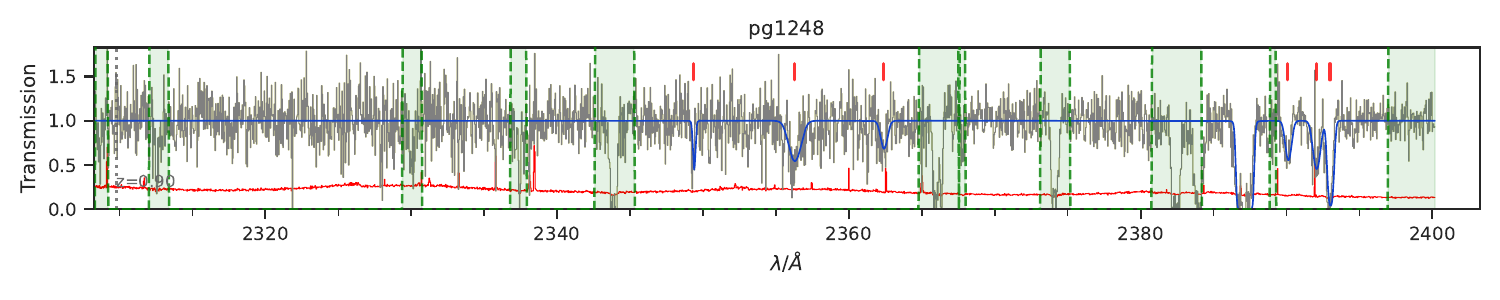}  
     {Fig.~\ref{fig:list_spe_1} continued. Spectrum of PG1248+401.}
     
  \label{fig:list_spe_1}
\end{figure*} 

 \begin{figure*}
\centering
    \includegraphics[width=0.95\linewidth]{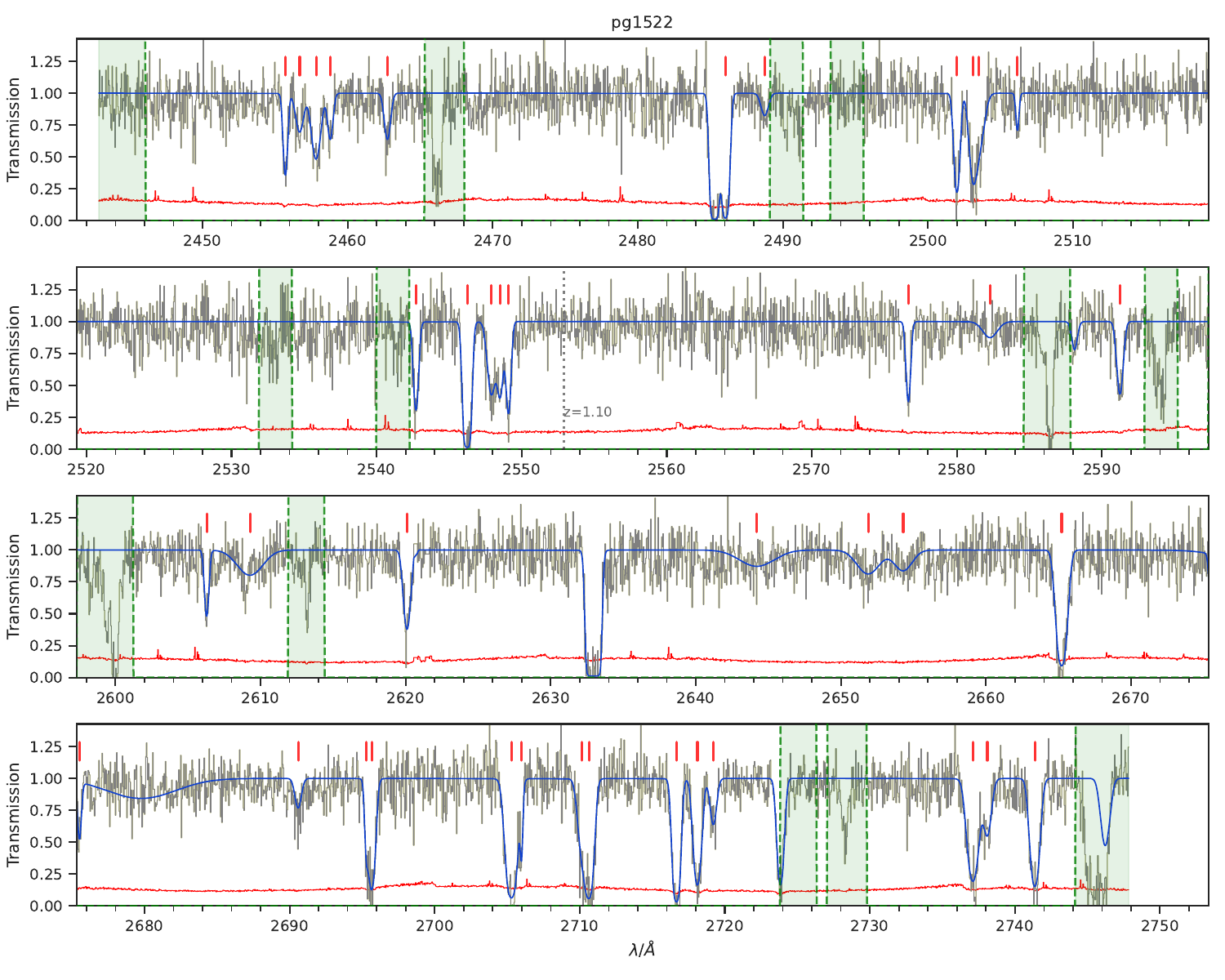}
    {Fig.~\ref{fig:list_spe_1} continued. Spectrum of PG1522+101.}
  \label{fig:list_spec_2}
\end{figure*}

 \begin{figure*}
\centering
    \includegraphics[width=0.95\linewidth]{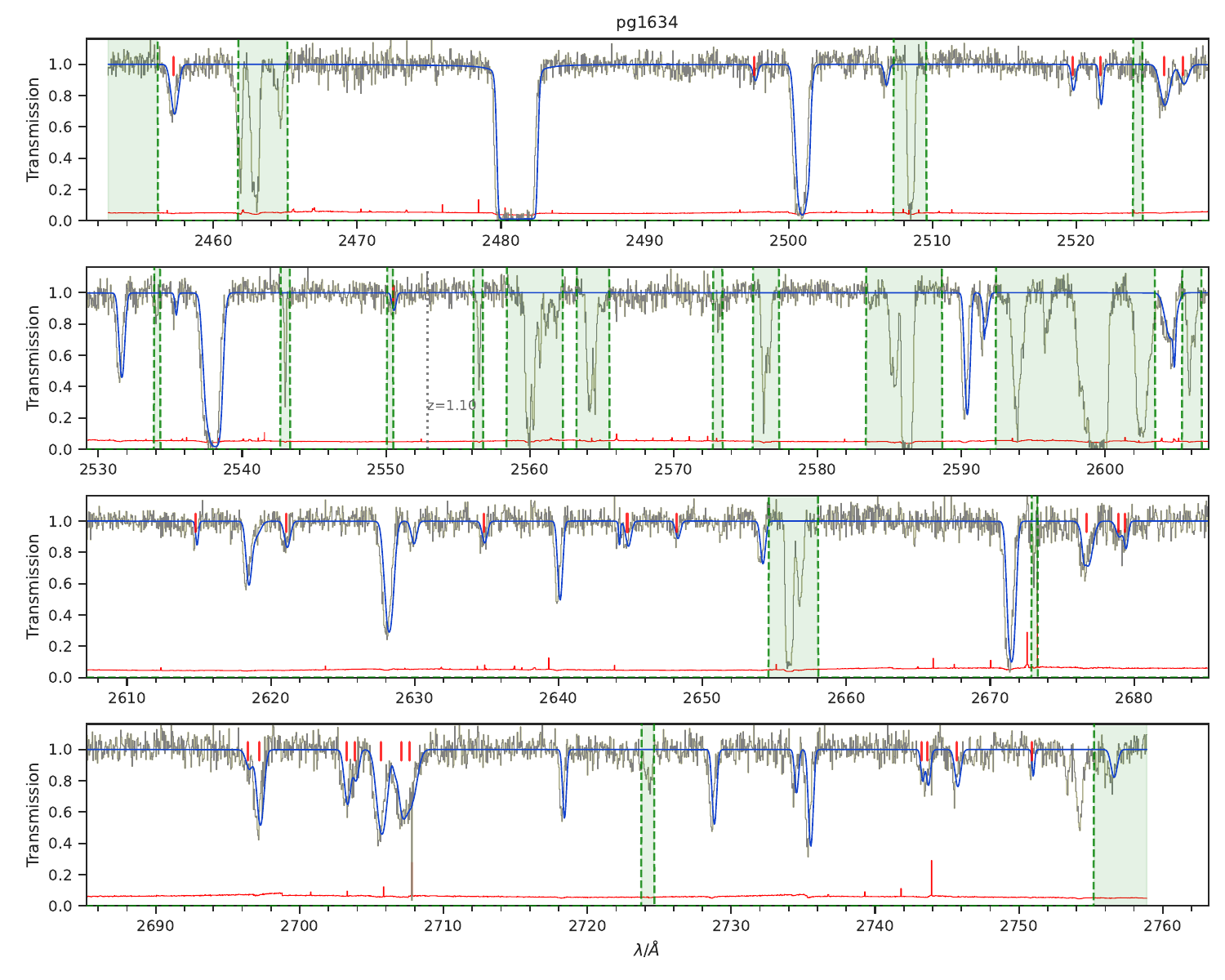}
    {Fig.~\ref{fig:list_spe_1} continued. Spectrum of PG1634+706.}
  \label{fig:list_spec_3}
\end{figure*} 

 \begin{figure*}
\centering
    \includegraphics[width=0.95\linewidth]{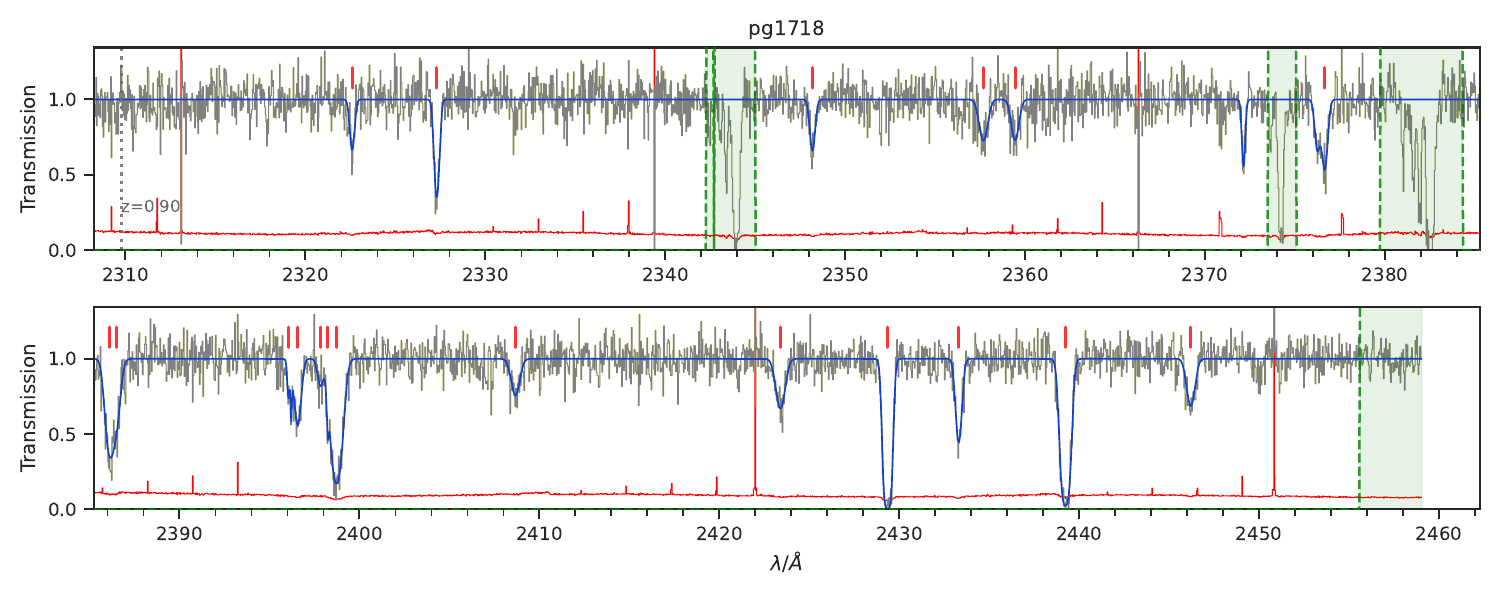}
    {Fig.~\ref{fig:list_spe_1} continued. Spectrum of PG1718+481.}
  \label{fig:list_spec_4}
\end{figure*}

 \begin{figure*}
\centering
    \includegraphics[width=0.95\linewidth]{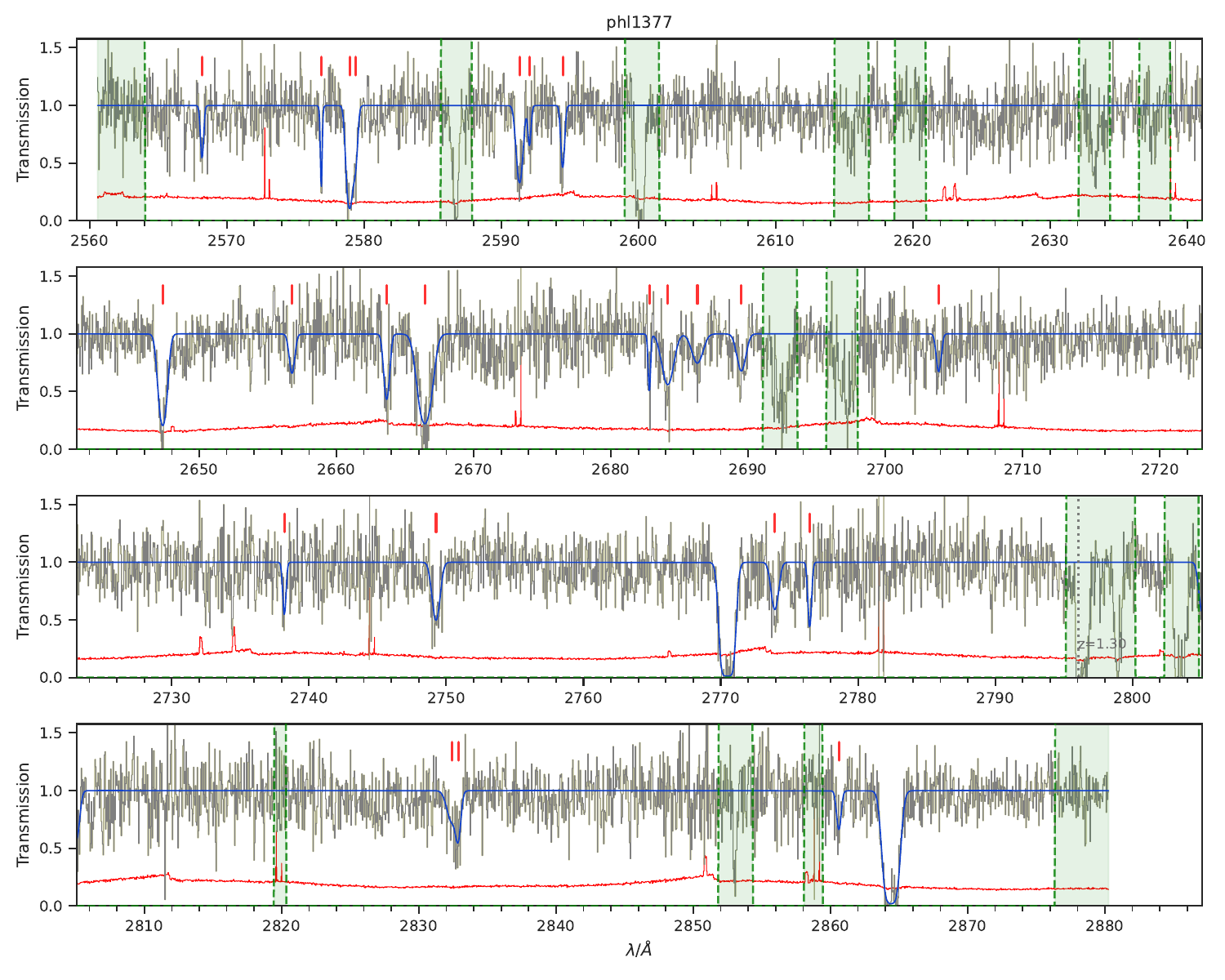}
    {Fig.~\ref{fig:list_spe_1} continued. Spectrum for PHL1377.}
    \includegraphics[width=0.95\linewidth]{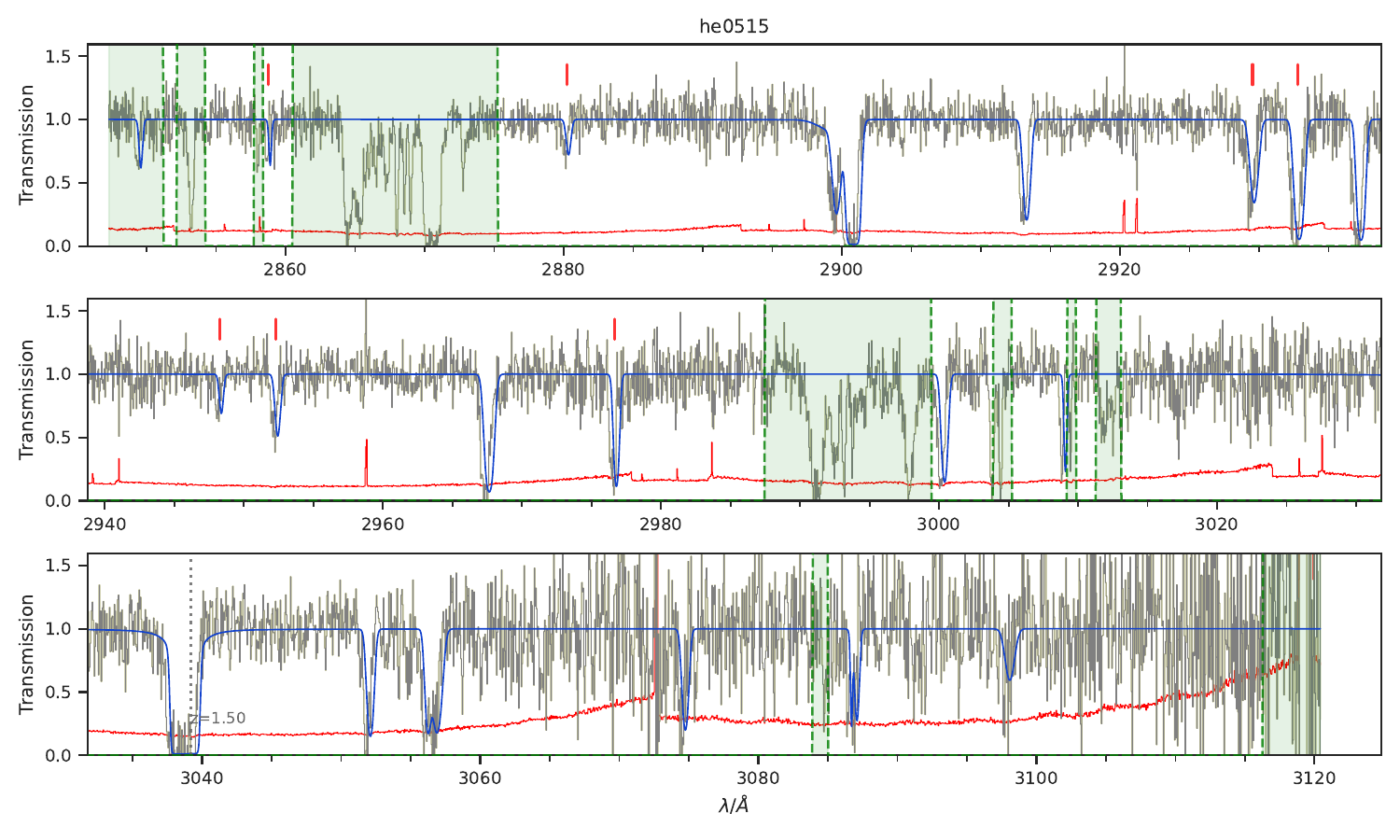}
    {Fig.~\ref{fig:list_spe_1} continued. Spectrum of HE0515-4414.}
  \label{fig:list_spec_5}
\end{figure*} 

 \begin{figure*}
\centering
    \includegraphics[width=0.95\linewidth]{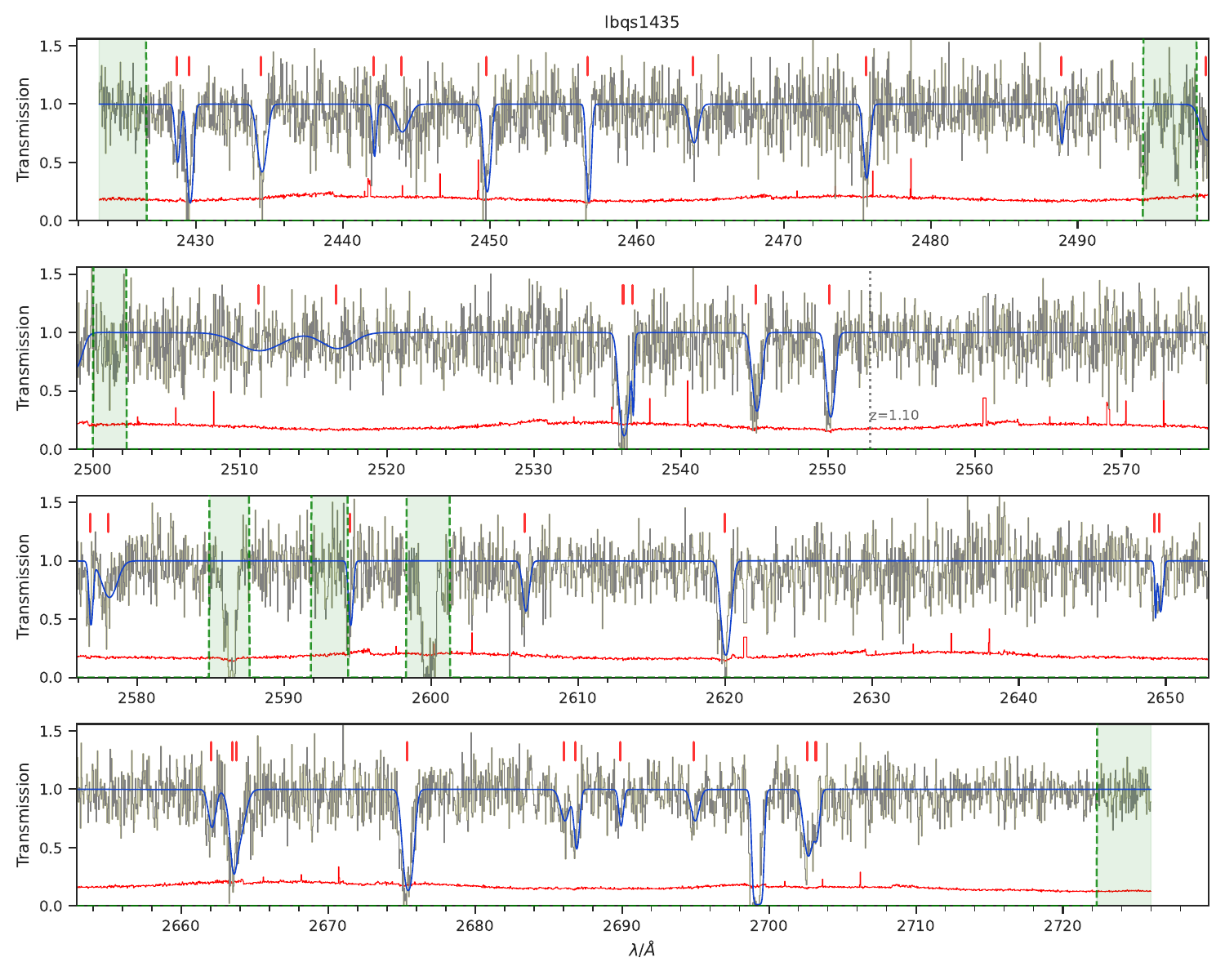}
    {Fig.~\ref{fig:list_spe_1} continued. Spectrum of LBQS1435-0134.}
  \label{fig:list_spec_6}
\end{figure*}

 \begin{figure*}
\centering
    \includegraphics[width=0.95\linewidth]{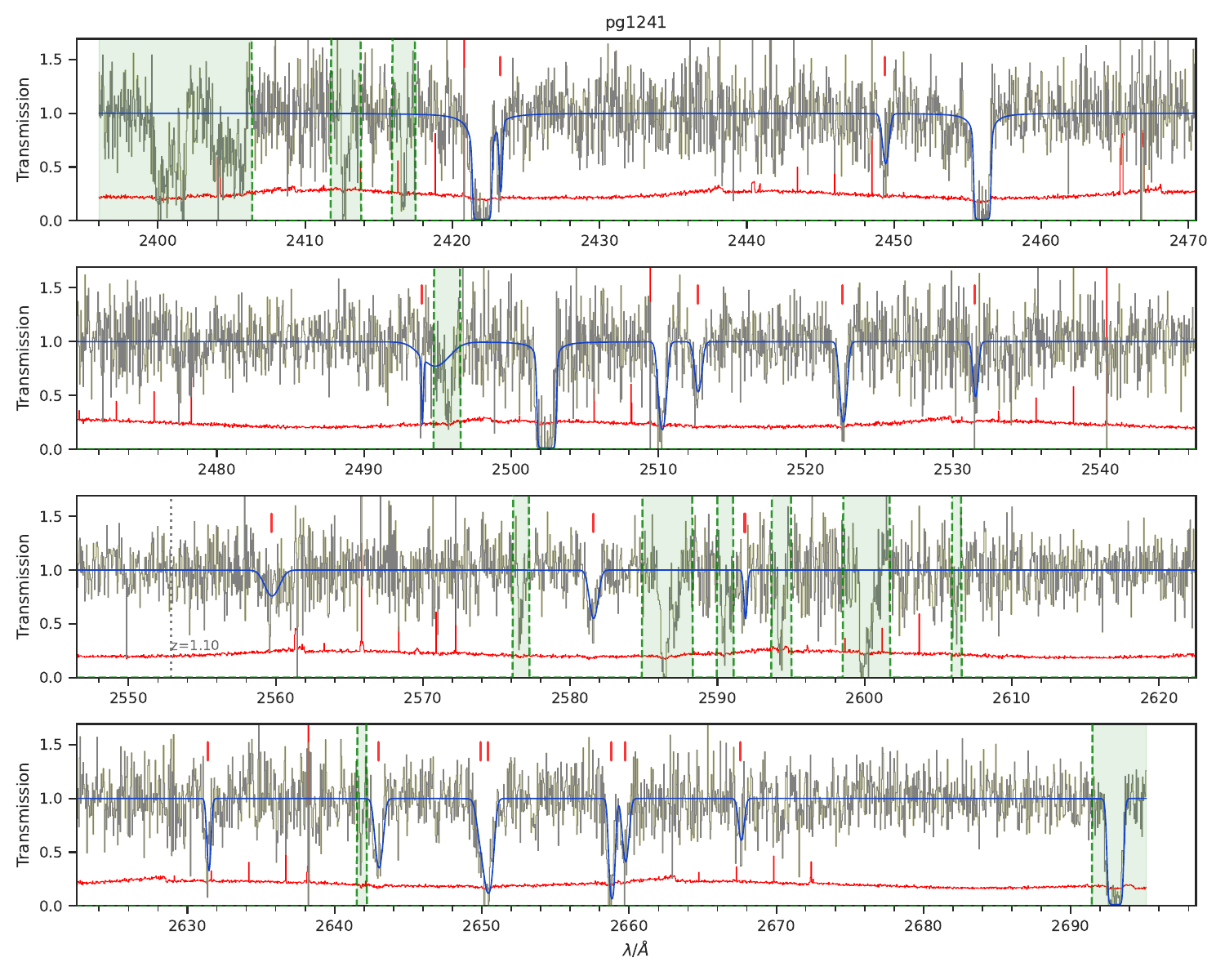}
    {Fig.~\ref{fig:list_spe_1} continued. Spectrum of PG1241+176.}
  \label{fig:list_spec_7}
\end{figure*} 

 \begin{figure*}
\centering
    \includegraphics[width=0.95\linewidth]{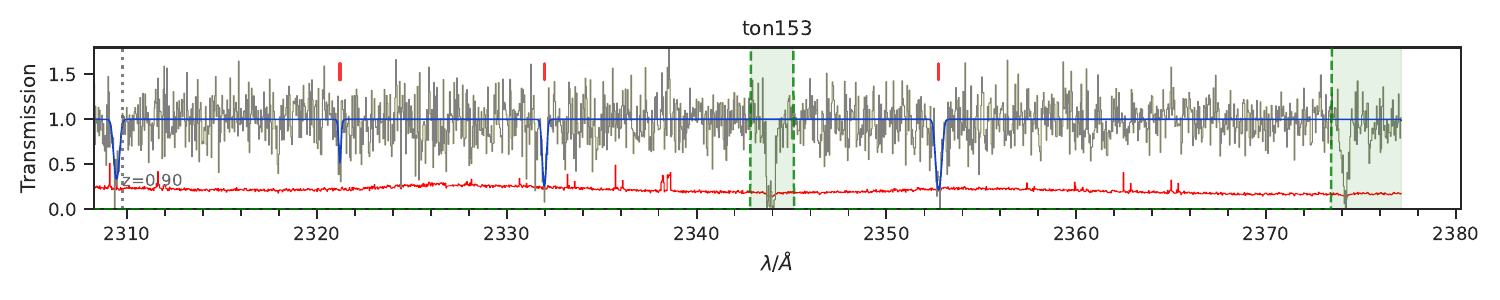}
    {Fig.~\ref{fig:list_spe_1} continued. Spectrum of TON153.}
  \label{fig:list_spec_8}
\end{figure*}

 \begin{figure*}
\centering
    \includegraphics[width=0.95\linewidth]{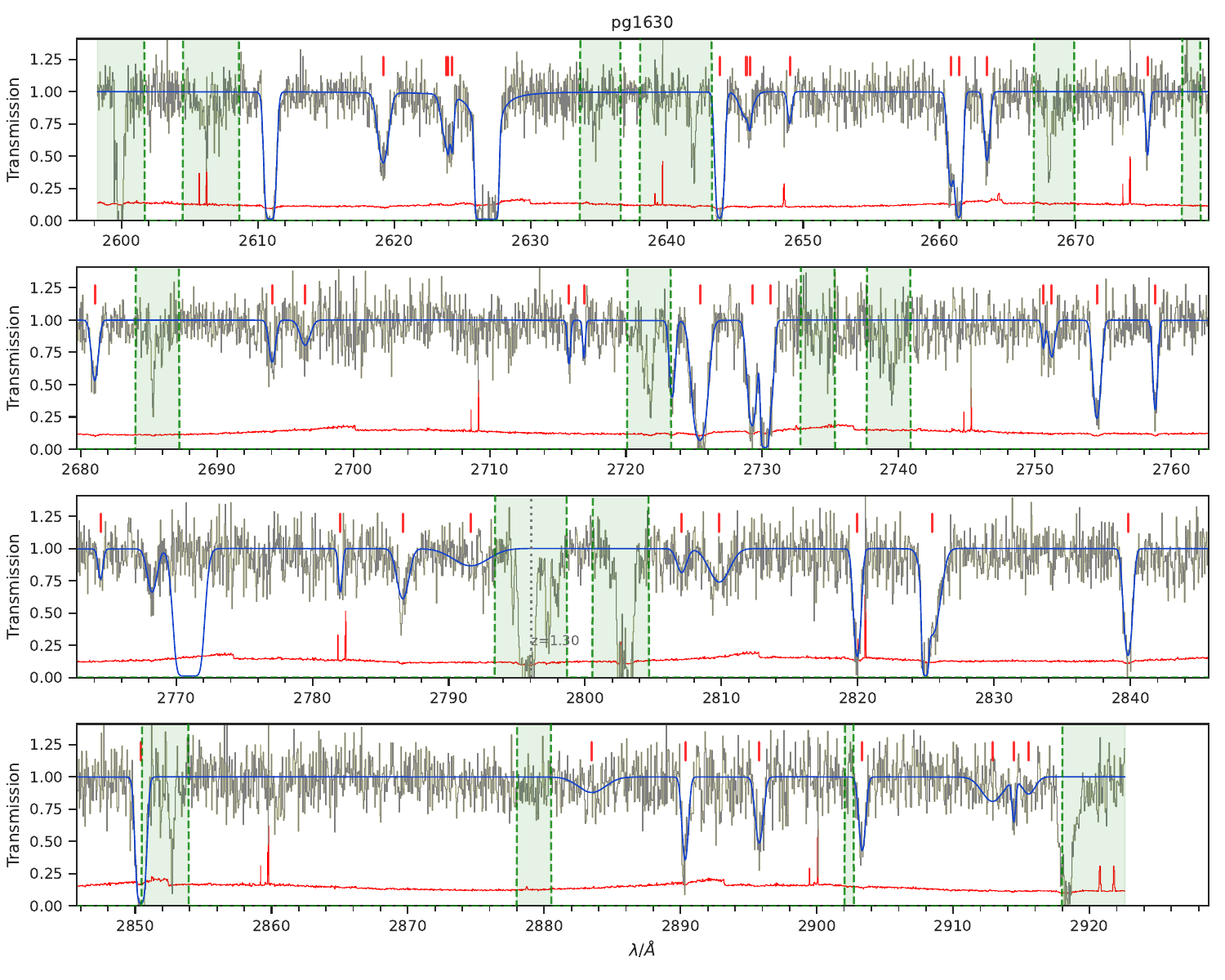}
    {Fig.~\ref{fig:list_spe_1} continued. Spectrum of PG1630+377.}
  \label{fig:list_spec_10}
\end{figure*} 

\bsp	
\label{lastpage}
\end{document}